\documentclass[preprint,showpacs,preprintnumbers,amsmath,amssymb,aps]{revtex4}

\usepackage{graphicx}       % Include figure files
\usepackage{dcolumn}        % Align table columns on decimal point
\usepackage{bm}             % bold math
\usepackage{psfrag}

\newcommand{\nn}{\nonumber}
\newcommand{\grad}{\nabla}

\newcommand{\lam}{\lambda}
\newcommand{\delsl}{\not\!\partial}
\newcommand{\gam}{\gamma}

\newcommand{\sig}{\sigma}

\newcommand{\half}{{\textstyle \frac{1}{2}}}

\newcommand{\eps}{\epsilon}

\newcommand{\ud}{\mathrm{d}}

\newcommand{\bp}{\mbox{\boldmath $p$}}
\newcommand{\bP}{\mbox{\boldmath $P$}}
\newcommand{\bk}{\mbox{\boldmath $k$}}
\newcommand{\bq}{\mbox{\boldmath $q$}}
\newcommand{\bs}{\mbox{\boldmath $s$}}
\newcommand{\bsig}{\mbox{\boldmath $\sigma$}}

\newcommand{\Mcal}{\mathcal{M}}

\newcommand{\psl}{\not\!p}
\newcommand{\ksl}{\not\!k}
\newcommand{\rsl}{\not\!R}
\newcommand{\ssl}{\not\!s}
\newcommand{\asl}{\not\!a}
\newcommand{\bsl}{\not\!b}
\newcommand{\csl}{\not\!c}
\newcommand{\dsl}{\not\!d}
\newcommand{\integrand}{\int \frac{d^3k}{(2\pi)^3}}

\newcommand{\halfa}{\tfrac{1}{2}}
\newcommand{\halfb}{\frac{1}{2}}
\newcommand{\halfc}{1/2}

\newcommand{\Aoutalt}{A_\kappa^\text{out}}
\newcommand{\Ainalt}{A_\kappa^\text{in}}

\newcommand{\rmin}{r_\text{min}}
\newcommand{\rmax}{r_\text{max}}
\newcommand{\bcrit}{b_c}
\newcommand{\bLhat}{\hat{\mathbf{L}}}
\newcommand{\polP}{\mathbf{\mathcal{P}}}
\newcommand{\mass}{m}

\newcommand{\us}{\overline{u}_s(\bp_f)}
\newcommand{\ur}{u_r(\bp_i)}
\newcommand{\sigabs}{\sigma_\text{A}}
\newcommand{\bQ}{\mathbf{Q}}
\newcommand{\br}{\mathbf{R}}

\newcommand{\Tr}{\text{Tr}}
\newcommand{\bspin}{C}

\begin{document}

\preprint{}

\title{Fermion Scattering by a Schwarzschild Black Hole}

\author{Sam Dolan}
 \email{s.dolan@mrao.cam.ac.uk}
\author{Chris Doran}
 \email{c.doran@mrao.cam.ac.uk}
\author{Anthony Lasenby}
 \email{a.n.lasenby@mrao.cam.ac.uk}
 \affiliation{%
 Cavendish Laboratory, University of Cambridge, J J Thomson Avenue, Cambridge CB3 0HE, UK\\
}%

\date{\today}

\begin{abstract}
We study the scattering of massive spin-half waves by a Schwarzschild
black hole using analytical and numerical methods.  We begin by
extending a recent perturbation theory calculation to next order to
obtain Born series for the differential cross section and Mott
polarization, valid at small couplings. We continue by deriving an
approximation for glory scattering of massive spinor particles by
considering classical timelike geodesics and spin precession. Next, we
formulate the Dirac equation on a black hole background, and outline a
simple numerical method for finding partial wave series
solutions. Finally, we present our numerical
calculations of absorption and scattering cross sections and
polarization, and compare with theoretical expectations.

\end{abstract}

\maketitle

\section{\label{sec:introduction}Introduction}
Scattering by gravitational sources has been an important test for
General Relativity since its inception. An early success for
Einstein's theory came with verification of the prediction that
starlight is deflected as it passes close to the Sun. The experiments
conducted by Eddington and his team during the solar eclipse of 1919
were important in establishing GR as a credible theory of
gravitation. More recently, gravitational lensing observations provide
a strong test of the theory, as well as information about lensing mass
distributions and distant astronomical sources.

Gravitational scattering from astrophysical objects may be understood
entirely in terms of geometric optics through the analysis of
classical geodesics. Nevertheless, to obtain a deeper theoretical
understanding of extreme objects such as black holes many authors have
also considered the scattering of coherent waves. If the wavelength of
the incident wave is comparable with the size of the event horizon
then the wave will be diffracted by the black hole. Diffraction
effects are responsible for many interesting phenomena in nature, such
as glories and rainbows, so wave scattering from black holes is an
interesting field in its own right, even if observations may not be
realisable in practice.

In this paper, we study how quantum waves are scattered by a
Schwarzschild black hole. This problem has been considered in numerous
papers \nocite{Matzner-1968, Fabbri-1975, Peters-1976, DeLogi-1977, Sanchez-1976, Sanchez-1977,
Sanchez-1978a, Sanchez-1978b, Zhang-1984, Matzner-1985, Anninos-1992,
Andersson-1995, Andersson-2001, Doran-2002}\cite{Matzner-1968}--\cite{Doran-2002} and books
\cite{Chandrasekhar-1983, Futterman-1988, Frolov-1998}. Most authors
study the massless scalar wave (with spin $s = 0$) for its
mathematical simplicity, or massless electromagnetic ($s = 1$) and
gravitational ($s = 2$) waves for physical relevance. In this paper,
we focus instead on fermion scattering ($s = 1/2$), which has been
less frequently discussed \cite{Futterman-1988}. We investigate the
effect of particle mass on scattering cross sections and polarization.

The interaction of an incident plane wave with a black hole may be
understood in terms of three quantities: the total absorption cross section
$\sig_A$, the differential scattering cross section
$d\sig / d\Omega$, and the polarization $\polP$ that is induced
in initially unpolarized beam (note that $\frac{d\sig}{d\Omega}$ and
$\polP$ are functions of the scattering angle $\theta$). In
this paper we calculate these quantities using a variety of
analytical and numerical methods.

The paper is organised as follows. In section \ref{sec:classical} we
review classical scattering on the Schwarzschild background. We derive
a new approximation for near-horizon deflection of timelike geodesics,
and investigate classical spin precession on backward-scattering
orbits. In section \ref{sec:analytic} we describe analytic methods
that can be applied to fermion scattering. The perturbation-theory
approach introduced by Doran and Lasenby \cite{Doran-2002} is taken to
the next order, which provides a new estimate of the cross section and
polarization at low couplings. The glory-scattering approximation of
Zhang and DeWitt-Morette \cite{Zhang-1984} is extended to the massive case, 
employing the classical results of section \ref{sec:classical}. In
section \ref{sec:spinorscattering} we discuss partial wave scattering
theory, formulate the Dirac equation, and outline a simple numerical
method for calculating phase shifts. In section \ref{sec:results} we
present the results of our numerical calculations, and compare with
theory. We conclude in section \ref{sec:discussion} by discussing the
significance of our results and possible future work.

Note that in this article we employ units in which the speed of
light ($c$), the gravitational constant ($G$), and Planck's constant
over $2\pi$ ($\hbar$) are set to unity.

\section{\label{sec:classical}Classical Scattering}
We begin by reviewing some results for classical scattering and
absorption on a Schwarzschild background. We then investigate the
precession of classical spin vectors along scattering trajectories. As well
as being of interest in their own right, the results of this section will aid our 
interpretation of the wave scattering cross sections presented in
section \ref{sec:results}. The results of this section are also
required in section \ref{sec:analytic} to derive a new approximation
for glory scattering of massive spinor waves.

The Schwarzschild spacetime is invariant under time-displacement and
rotation. These symmetries imply the existence of Killing vectors, and
the quantities conjugate to those Killing vectors are conserved along
geodesics. That is, the particle energy $E$, and angular momentum $L$,
are constants of motion.
%, defined by
%\begin{equation}
%(1 - 2M/r) \frac{dt}{d\tau} \equiv E/\mass , \quad \quad  r^2 \frac{d\phi}{d\tau} \equiv L/\mass , \nn
%\end{equation}
%are constants of motion. Here, $\{t, r, \theta, \phi\}$ are the standard coordinates of the
%Schwarschild metric, $\tau$ is proper time, and $\mass$ is the mass of the particle.

%A manifest disadvantage of the Schwarzschild time coordinate $t$ is that it is
%only valid in the exterior region $r > 2M$; it takes an infinite
%coordinate time for ingoing geodesics to reach the horizon. In Section
%\ref{sec:spinorscattering} we discuss alternative coordinate systems
%with non-diagonal line elements that are valid from $r > 0$.

The motion of a particle of rest mass $\mass$ on a Schwarzschild background
is described by the orbital equation
\begin{equation}
\left(\frac{du}{d\phi}\right)^2 = 2Mu^3 - u^2 + \frac{2Mm^2}{L^2}u + \frac{E^2 - \mass^2}{L^2}
\label{classical-de-first},
\end{equation}
where $u = 1 / r$, and $\phi$ is the polar angle. For descriptive
purposes it is useful to define three further constants: the
momentum $p = (E^2 - \mass^2)^{1/2}$, the impact parameter $b = L
/ p$, and the speed $v = p / E$. At large distances from the
hole, each has a simple physical interpretation: $p$ is the incident
particle momentum, $b$ is the orthogonal distance to the scattering
centre, and $v$ is the particle speed in units of $c$. Note that
for null geodesics, $v = 1$ and $p = E$. With these definitions we may
rewrite the orbital equation (\ref{classical-de-first}) as
%In the weak-field limit,
%\begin{equation}
%\frac{1}{r^4}\left(\frac{d r}{d\phi}\right)^2 = \frac{1}{b^2} + \frac{2Mm^2}{b^2p^2r} - \frac{1}{r^2}\left[1 - \frac{2M}{r} \right] 
%\label{geodesicequation}
%\end{equation}
\begin{equation}
\left( \frac{d u}{d \phi} \right)^2 = 2M u^3 - u^2 + \frac{2M (1 - v^2)}{v^2 b^2} u + \frac{1}{b^2} 
\label{classical-de}.
\end{equation}
The unbound scattering solutions of (\ref{classical-de}) can be
written in terms of elliptic integrals.  For a comprehensive
treatment, see Chandrasekhar \cite{Chandrasekhar-1983}. %Below we
%consider two useful
%approximations that are valid in the weak-field limit and near-horizon limit.
\subsection{Deflection-angle Approximations}
\begin{figure}
\begin{center}
\includegraphics[width=6cm]{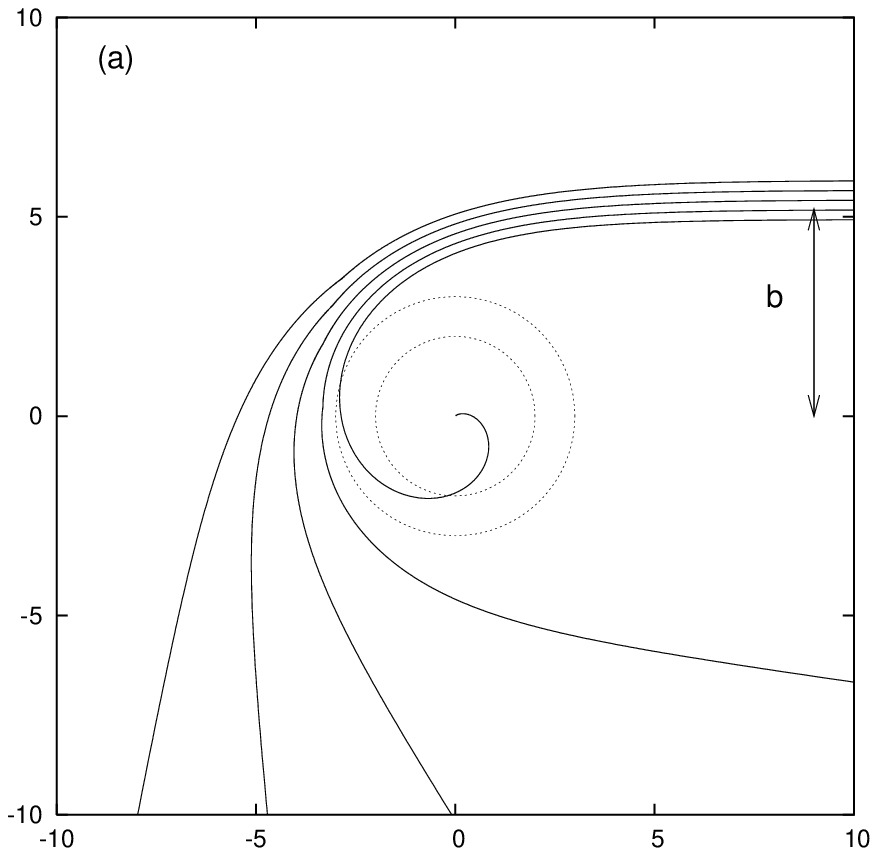}
\includegraphics[width=8.6cm]{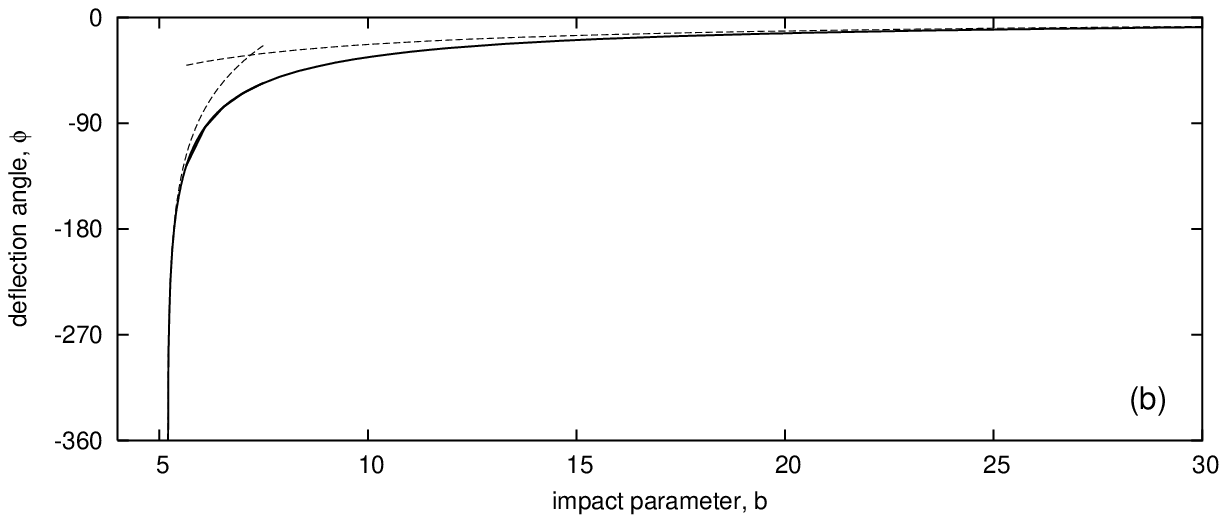}
\end{center}
\caption[dummy1]{(a) Photon geodesics around a Schwarzschild
black hole, $M = 1$, close to the critical impact parameter $\bcrit =
3\sqrt{3}M \approx 5.196M$. The inner circle shows the event horizon
at $r=2M$, and the outer circle shows the unstable photon orbit at
$r=3M$. (b) Deflection angle $\phi$ as a function of impact
parameter $b$. The plot compares the scattering angle with (i) the
Einstein deflection angle $4M / b$, valid when $b \gg \bcrit$, and
(ii) Darwin's approximation, $b \approx 3 \sqrt{3} M + 3.48 M 
e^{-\phi}$, valid when $b \sim \bcrit$.}
\label{spiral}
\end{figure}

In the weak-field limit, $r \gg 2M$, the deflection angle for timelike geodesics is approximately
\begin{equation}
\Delta \phi \approx \frac{2M (1 + v^2)}{b v^2}.
\label{einstein-approx}
\end{equation}
For null geodesics we set $v = 1$ to recover the Einstein deflection
angle for the bending of light, $\Delta \phi \approx 4M / b$. In the
strong-field regime, the above approximation 
underestimates the deflection; incoming geodesics may be scattered through
large angles or even orbit the hole several times before escaping (see
Fig.  \ref{spiral}). Trajectories that make too close an approach to the hole
spiral inwards, and end on the singularity. The last geodesic to avoid the
singularity defines a critical impact parameter $\bcrit$. All
geodesics with $b > \bcrit$ are scattered, whereas all geodesics with
$b < \bcrit$ are absorbed. By considering the zeros of
(\ref{classical-de}), one can show (see
e.g. \cite{Unruh-1976-absorption}) that
\begin{equation}
\bcrit = \frac{M}{\sqrt{2} v^2} \left(8 v^4 + 20v^2 - 1 + (1 + 8 v^2)^{3/2} \right)^{1/2}
\label{classicalbcrit}.
\end{equation}
For null geodesics, the critical impact parameter
is $\bcrit = 3\sqrt{3}M$. The classical absorption cross section is
simply the area of the circle defined by the critical impact
parameter, $\sigma_A = \pi {\bcrit}^2$. 

Many years ago, Darwin \cite{Darwin-1959} derived an approximation for
the deflection of null (massless) rays that pass close to the critical
orbit. He found that the deflection angle $\Theta$ is
related to the impact parameter $b$ by the equation
\begin{equation}
b - \bcrit =  216 \times 3 \sqrt{3} M \left( \frac{\sqrt{3} - 1}{\sqrt{3} + 1} \right)^2 e^{-\pi} e^{-\Theta} \approx 3.48 M e^{-\Theta} .
\label{darwin1}
\end{equation} 
Using the same techniques, an approximation for the deflection of
timelike (massive) geodesics can be derived. This calculation is outlined in
Appendix A, and the result is
\begin{equation}
b - b_c = M \frac{{2}^{13/2} e_c^3 (e_c+3)^{3/2}}{v (e_c + 1)^{5/2}} \left( \frac{ 1 - \sqrt{ (e_c - 1) / 2e_c } } { 1+  \sqrt{ (e_c - 1) / 2e_c }} \right)^2 \exp \left(- \pi \sqrt{\frac{2e_c}{e_c+3}} \right) \exp \left( - \sqrt{\frac{2e_c}{e_c+3}} \Theta \right) 
\label{darwin2},
\end{equation}
where $e_c = (1 + 8v^2)^{1/2}$. This result reduces to equation
(\ref{darwin1}) in the case $v = 1$.

\begin{figure}
\begin{center}
\includegraphics[width=8.6cm]{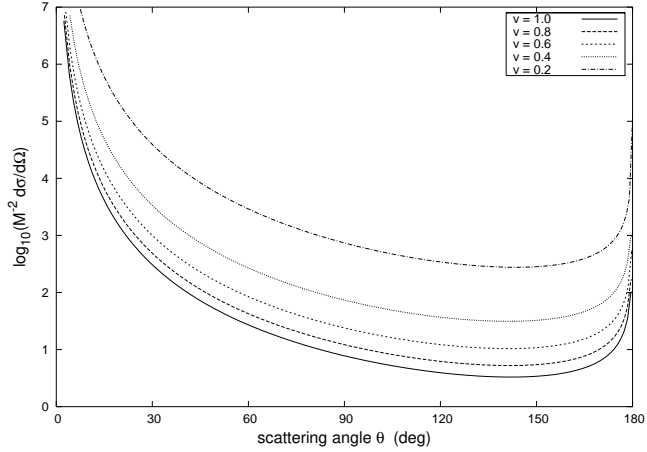}
\end{center}
\caption[dummy1]{Classical scattering cross sections for a range of particle speeds. The cross sections diverge on-axis at $\theta = 0$ and $\theta = \pi$ as $\frac{d\sigma}{d\Omega} \approx 4 v^{-4} (1+v^2)^2 M^2 \theta^{-4}$ and $\frac{d\sigma}{d\Omega} \sim |\pi - \theta|^{-1}$ respectively. }
\label{classical-scattering-cross-section}
\end{figure}

A classical scattering cross section may be derived by
considering a stream of parallel incident geodesics, as shown by
Collins, Delbourgo and Williams \cite{Collins-1973}. Geodesics passing close to
the critical orbit are scattered through angles that may be many
multiples of $2 \pi$, so the total cross section is an infinite sum,
\begin{equation}
\frac{d \sigma}{d \Omega} = \frac{1}{\sin \theta} \sum{ b(\theta) \left| \frac{db}{d\theta} \right| }, 
\end{equation}
where the sum is taken over angles $\theta, 2 \pi - \theta, 2 \pi +
\theta, 4\pi - \theta$, etc. The cross section is divergent in the
forward direction, $d\sigma / d\Omega \approx 4 v^{-4} (1+v^2)^2 M^2
\theta^{-4}$ \cite{Collins-1973}, due to the long-range nature of the interaction. It is
also divergent (but integrable) in the backward direction,
$d\sigma / d\Omega \propto |\pi - \theta|^{-1}$, because the
element of solid angle tends to zero on-axis. The behaviour of the
cross section close to $\theta = 0$ and $\theta = \pi$ may be
understood using the approximations (\ref{einstein-approx}) and
(\ref{darwin2}) given above, and at intermediate angles it may be
found by numerically integrating the elliptic equations. The result is
plotted in Fig. \ref{classical-scattering-cross-section} for a range
of particle speeds.

\subsection{\label{sec:classical:spin}Spin Precession}
In the next section (\ref{sec:analytic}) we derive a
semi-classical approximation to the massive fermion scattering cross
section in the backward direction. To do
this, we must first consider how the spin of the particle is affected
by the gravitational interaction. For a massless wave, this is
straightforward, as the helicity of the Dirac spinor remains
unchanged. For a massive wave, the situation is substantially more
complicated. However, progress can be made by considering the
precession of the classical spin vector. We can then apply a 
semi-classical argument to relate the classical spin vector to the
Dirac spinor itself in the $\lambda \ll r_s$ limit.

We begin by introducing an angular momentum four-vector $s^\mu$ with
components $[s^t, s^r, s^\theta, s^\phi]$, where $t,r,\theta,\phi$ are
the coordinates of the standard Schwarzschild metric $g_{\mu \nu}$. The
four-vector $s^\mu$ is spacelike, normalised, and orthogonal to the velocity
$v^\mu$. Classically, angular momentum is parallel-transported along
the geodesic, that is, $\dot{s}^\mu + {\Gamma^\mu}_{\nu \lambda} v^\nu s^\lambda =
0$. Inserting the standard Schwarzschild connection coefficients, and
applying the orthogonality condition, we find the precession equations
can be written as a pair of coupled first order equations
\begin{equation}
\frac{d{\bar{s}}^\phi}{d \phi}  + s^r = 0 , \quad
\frac{d{s}^r}{d \phi} - (1 - 3M/r) \bar{s}^\phi = 0
\label{spinprecession},
\end{equation}
where $\bar{s}^\phi = r s^\phi$. Alternatively they can be combined in
a simple second-order equation
\begin{equation}
\frac{d^2 \bar{s}^\phi}{d \phi^2} + \left(1 - \frac{3M}{r} \right) \bar{s}^\phi = 0. 
\end{equation}

Now let us investigate a specific case, in which the spin vector is
initially aligned with the direction of motion. At infinity, where $u
= 0$, we start with initial conditions $s^r = -\gamma$ and
$\bar{s}^\phi = 0$, where $\gamma = (1-v^2)^{-1/2}$. The solutions to
the precession equations may then be written as
\begin{align}
 \bar{s}^\phi(u) =& \left(1 + u^2 / \eta^2 \right)^{1/2} \sin \chi(u)  \nn \\
 s^r (u)   =& -\gamma \left(1 + u^2 / \eta^2 \right)^{-1/2} \cos \chi(u) \nn \\
                 & - (u / \eta^2)  \left(1 + u^2 / \eta^2 \right)^{-1/2} \left( 2Mu^3 - u^2 + 2M \eta^2 u + 1/b^2 \right)^{1/2} \sin \chi(u) , 
\end{align}
where $\eta = (\gamma b v)^{-1}$ and 
\begin{equation}
\chi(u) = \gamma \int_{0}^u \left(1 + w^2 / \eta^2 \right)^{-1} \left( 2Mw^3 - w^2 + 2M \eta^2 w + 1/b^2 \right)^{-1/2} dw. 
\label{chi}
\end{equation}
For comparison, the scattering angle can be written in similar form as
\begin{equation}
\phi(u) = \int_{0}^u \left( 2Mw^3 - w^2 + 2M \eta^2 w + 1/b^2 \right)^{-1/2} dw. 
\label{phi}
\end{equation}
Equations (\ref{chi}) and (\ref{phi}) can be expressed in terms of elliptic functions, but we
prefer to solve them numerically here.
%Figure \ref{precession-orbit} illustrates how the direction of spin vector changes around a backward-scattering orbit, as seen in the Schwarzschild frame.
%\begin{figure}
%\begin{center}
%\includegraphics[width=12cm]{precession_multiplot.eps}
%\end{center}
%\caption[dummy1]{\emph{Spin precession around backward-scattering orbits}.}
%\label{precession-orbit}
%\end{figure}

\begin{figure}
\begin{center}
\psfrag{pi}{$\mathbf{p}_i$}
\psfrag{pf}{$\mathbf{p}_f$}
\psfrag{si}{$\mathbf{s}_i$}
\psfrag{sf}{$\mathbf{s}_f$}
\psfrag{chi}{$2 \chi_\infty$}
\psfrag{xi}{$\xi$}
\includegraphics[width=5cm]{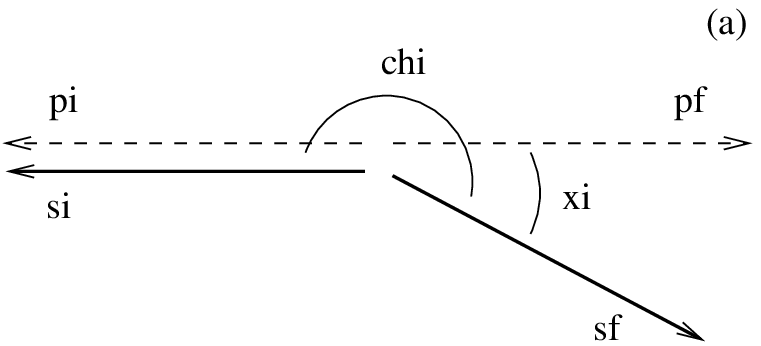} \\
\includegraphics[width=8.6cm]{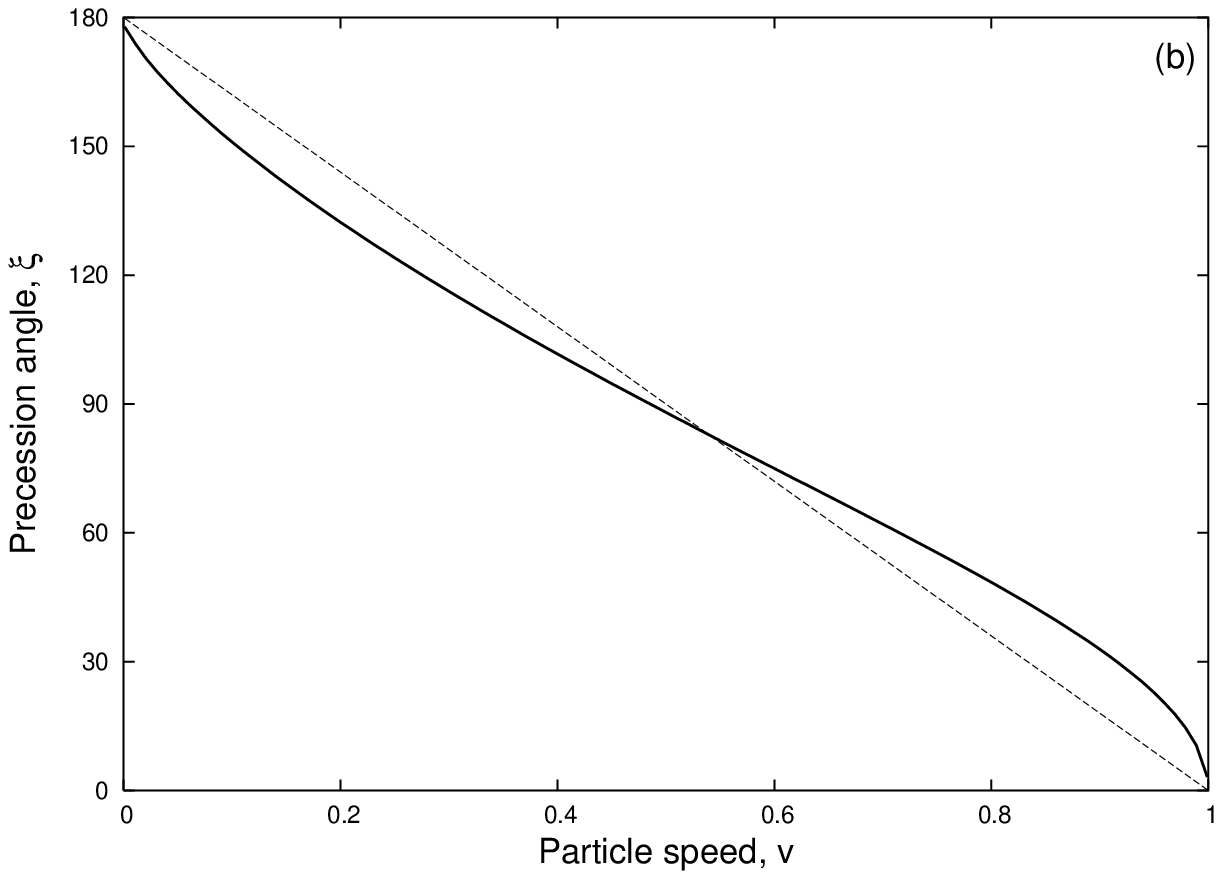}
\end{center}
\caption[dummy1]{Spin precession around a $180^\circ$ orbit. (a) Depicting the precession angle $\xi$. Here, $\mathbf{p}_i$ and $\mathbf{p}_f$ are the initial and final momentum directions, and $\mathbf{s}_i$ and $\mathbf{s}_f$ are the initial and final spin directions in the rest frame of the particle. (b) The precession angle $\xi$ (solid line) as a function of particle velocity $v$. The dotted line is for reference only.}
\label{chi-fig}
\end{figure}

In the next section we discuss glory scattering, which is related
to the classical orbits that emerge in the backwards direction. We are
particularly interested in the angle $\xi$ that the classical spin
vector makes with the outgoing radial direction, in the rest frame of the
particle. For a clockwise orbit, we define
\begin{equation}
\xi \equiv \lim_{r \rightarrow \infty} \left( \tan^{-1} \left(\frac{-\bar{s}^\phi}{s^r / \gamma}\right) \right) = 2 \chi_\infty - \pi ,
\label{xi}
\end{equation}
where $\chi_\infty = \chi(u_{0})$, and $u_{0} = 1 / r_{0}$ is found from the radius of closest approach for the $180^\circ$ orbit. The vectors and angles in Eq. (\ref{xi}) are illustrated in Fig. \ref{chi-fig}(a). Figure \ref{chi-fig}(b) plots $\xi$ for the backward-scattering orbit as a function of particle velocity, and demonstrates that $\xi$ decreases monotonically as $v$ increases, from $\xi = \pi$ at $v = 0$, to $\xi = 0$ at $v = 1$.

\section{\label{sec:analytic}Spinor Scattering: Analytic Results}
In this section we review some approaches to calculating the fermionic
cross sections for black hole scattering. The appropriate method 
depends on the ratio of the black hole size to particle wavelength. A
convenient dimensionless measure of the gravitational coupling is
\begin{equation}
\eps = \frac{G M E}{\hbar c^3} = \frac{\pi r_S}{\lambda v}
\end{equation}
where $r_S$ is the Schwarzschild radius, and $\lambda = h / p$ is the
wavelength of the quantum particle. Reverting to natural units, we see
that $\eps = ME$, which we use to label our figures.

\subsection{Low-coupling scattering cross sections: Perturbation theory}
In a recent paper \cite{Doran-2002}, Doran and Lasenby showed that the
fermion scattering cross section can be expanded as a perturbation
series. In effect, they calculated the gravitational analogue of the
Mott formula for scattering from a Coulomb potential. The series takes the form
\begin{equation}
\frac{d \sigma}{d \Omega} = \left(\frac{GM}{c^2}\right)^2 \left[ a_0(v, \theta) + \eps a_1(v, \theta) + \eps^2 a_2(v, \theta) + \ldots \right],
\label{born-series}
\end{equation}
with $\eps$ as defined above, and $a_i(v, \theta)$ dimensionless
functions. The perturbation series approach is most appropriate in the
long-wavelength limit, when $\eps \ll 1$.  Doran and Lasenby showed
that the first-order term $a_0$ is gauge-invariant and equal to
\begin{equation}
a_0(v, \theta) = \frac{ 1 + 2v^2 - 3v^2 \sin^2(\theta / 2) + v^4 - v^4 \sin^2(\theta/2)  }{ 4 v^4 \sin^4 (\theta / 2) } .
\label{first-order-scat}
\end{equation}
In the massless limit ($v \rightarrow 1$), this reduces to the
unpolarized Mott scattering formula. The cross section will tend to
the first-order result (\ref{first-order-scat}) in the long wavelength
limit, as $r_S / \lambda \rightarrow 0$. In Appendix B we show how to
calculate the next term in the perturbation series in the Kerr-Schild
gauge. We find
\begin{equation}
a_1(v, \theta) = \frac{\pi \left[(3+4v^2 + v^4)(2 - |\sin(\theta / 2)|) - (7v^2 + v^4) \sin^2(\theta/2)  \right]}{4 v^3 |\sin (\theta / 2)|^3} .
\label{second-order-scat}
\end{equation}
The equivalent result for Coulomb scattering was successfully
calculated by Dalitz \cite{Dalitz-1951} many years ago. The second
order matrix element also allows the calculation of the degree of
polarization $\polP$ induced in an initially unpolarized beam. The
net polarization is in the direction orthogonal to the plane of
scattering, $\bp_i \times \bp_f$.  To first order,
\begin{equation}
\polP = \left(\frac{G M \mass}{\hbar c} \right) \frac{v (3 + v^2) \tan^2(\theta / 2) \ln(\sin^2(\theta / 2))}{2 \left[1 + 2v^2 - 3v^2\sin^2(\theta / 2) + v^4 - v^4 \sin^2(\theta/2) \right]} \sin \theta
\label{second-order-pol}
\end{equation}
where we have included the dimensional constants for clarity. 
The confirmation of the gauge-invariance of results 
(\ref{second-order-scat}) and (\ref{second-order-pol}) awaits further
work. In section \ref{sec:results} we compare the perturbation series
and polarization with numerical results.

\subsection{Higher-coupling scattering cross sections: Glory and spiral scattering}
Many authors \cite{Sanchez-1978b, Matzner-1985, Anninos-1992} have noted
that black hole wave scattering produces diffraction effects that are
familiar from optical phenomena. Two such effects are glory and
spiral scattering. A glory is a bright spot or halo that appears
on-axis in the backward direction from the scatterer. Spiral
scattering, or orbiting, creates oscillations in scattered intensity
at intermediate scattering angles. In a general analysis of
semi-classical scattering, Ford and Wheeler \cite{Ford-1959}
showed that both effects may be understood with reference to the
deflection of classical paths. A glory occurs whenever the deflection
angle passes through a multiple of $\pi$, and spiral scattering occurs
when paths orbit the scattering centre multiple times. For a
discussion of the application of semi-classical techniques to the
black hole case, see \cite{Anninos-1992} or \cite{Futterman-1988}.

%gravitational field  key predictions of General Relativity is that 
%Eddington's measurement of the deflection of starlight by the Sun provided some of the earliest experimental support for Einstein's theory. 

%The Schwarzschild solution describes the space time exterior to a spherically-symmetric gravitational body 

%The scattering angle the Einstein angle, $\delta \theta = 4M / b$, 

%Quantum scattering. Can it be applied to black holes? Model system.
%Scattering from Black Holes, Futterman, Handler and Matzner \cite{Futterman-1988}.
%Chandrasekhar's book \cite{Chandrasekhar-1983}.
%Spin-half. Polarisation. Mott \cite{Mott-1932}. Mott and Massey \cite{Mott-1965}. Rose \cite{Rose-1961}.
%Matzner paper, \cite{Matzner-1968}.
%Sanchez Papers, \cite{Sanchez-1976, Sanchez-1977, Sanchez-1978a, Sanchez-1978b}.
%Andersson's phase-integral approximation \cite{Andersson-1995}.
%Glory scattering: Matzner et al. \cite{Matzner-1985}, Anninos et al. \cite{Anninos-1992}. 

Ford and Wheeler \cite{Ford-1959} derived a semi-classical
approximation of the glory scattering cross section for scalar ($s =
0$) waves. This approximation was extended to arbitrary spins by Zhang
and DeWitt-Morette in \cite{Zhang-1984} using path integral
methods. For massless waves, the backward glory cross section is
approximated by
\begin{equation}
\left. \frac{d \sigma}{d \Omega} \right|_\text{glory} \approx 2 \pi E {b_g}^2 \left|\frac{d b}{d \theta}\right|_{\theta = \pi} {J_{2s}}^2 (E b_g \sin \theta), 
\end{equation}
where $s$ is the spin of the particle, $J_{2s}$ is a Bessel function,
and $b_g$ is the impact parameter at which $\theta = \pi$. In
combination with Darwin's result (\ref{darwin1}) for the deflection of
a massless particle that passes close to the horizon, we have $b_g
\approx 5.3465 M$, and $|db / d\theta| = 0.1504 M$, so
\begin{equation}
M^{-2} \left. \frac{d \sigma}{d \Omega} \right|_\text{glory} \approx 2\pi EM \times 4.30 \times J^2_{2s} ( 5.3465 E M \sin \theta ).
\label{glory-approximation}
\end{equation}
For the scalar wave, the intensity has a peak in the backwards
direction, whereas for the spinor wave the intensity is zero
on-axis. 

We now derive a similar approximation for the massive spin-half case,
by taking into account the effects of classical spin precession. Let
us begin by considering a backward-scattering orbit defined by
$\mathbf{p}_i = \hat{z} = -\mathbf{p}_f$, that lies in the $\phi z$
plane. As we saw in Section \ref{sec:classical:spin}, the classical
spin vector is rotated by an angle of $2 \chi_\infty$ around this
orbit. The quantum spin vector, on the other hand, is found by the
double-sided action of the Dirac spinor, $s^\mu = \langle \psi |
\hat{\gam}^\mu \hat{\gam}_5 | \psi \rangle$, where $\hat{\gam}^\mu$ are
the standard Dirac-Pauli matrices. Assuming that the classical and
quantum spin vectors coincide in the classical limit, $\lambda / r_s
\rightarrow 0$, we conclude that, in the rest frame, the Dirac spinor
must transform according to
\begin{equation}
\psi_f \propto \left(  \cos (\chi_\infty) \hat{I} + \sin (\chi_\infty)  \hat{\gam}_\phi \hat{\gam}_3  \right) \psi_i
\label{R-explicit}
\end{equation}
where $\hat{I}$ is the identity matrix, and $\hat{\gam}_\phi = -\sin \phi
\hat{\gam}_1 + \cos \phi \hat{\gam}_2 $. Now consider the cross
section on-axis. Following the arguments of Zhang and DeWitt-Morette
\cite{Zhang-1984}, we note that there is a circular degeneracy here, and
that paths from all scattering planes will interfere
coherently. Around the circle $0 \le \phi < 2\pi$, the $\hat{\gam}_\phi
\hat{\gam}_3$ term of (\ref{R-explicit}) sums to zero, and the
intensity on-axis is therefore proportional to $\cos^2 (\chi_\infty)$.  Since it
is the spinors, rather than the vectors, that are summed coherently,
the half-angle of precession arises naturally. The
arguments of Zhang and DeWitt-Morette can be extended in a
straightforward way to derive an approximation for angles close to
$\pi$, leading to
\begin{align}
\left. \frac{d \sigma}{d \Omega} \right|_\text{glory} \approx 
2 \pi E v b^2 \left| \frac{db}{d\theta} \right| 
\left[  \begin{array}{ll}  & \cos^2 ( \chi_\infty ) {J_0}^2 (E b v \sin \theta) \\ + & \sin^2 ( \chi_\infty ) {J_1}^2 (E b v \sin \theta) \end{array} \right]
\label{glory-approx-massive-eq}.
\end{align}
Finally, we may use (\ref{darwin2}) to evaluate $b$ at $\Theta = \pi$,
and substitute into the above result. Alternatively, we may choose to
evaluate $b_g$ numerically for a more accurate result. In section \ref{sec:results} we compare both approaches with our numerical results.

\section{\label{sec:spinorscattering}Spinor Wave Scattering}
Before considering the Dirac equation on a Schwarzschild black
hole background, we first discuss the general theory of relativistic
spin-half scattering from a spherically-symmetric source.

\subsection{Scattering from a spherically-symmetric source}
In considering spherically-symmetric scattering we will employ the
two-component spherical spinors $\chi_\kappa^\mu$ that are
eigenvectors of the angular equation
\begin{equation}
(\bsig \cdot \bLhat + 1) \chi_\kappa^\mu = \kappa \chi_\kappa^\mu ,
\end{equation}
where $\bLhat = [\hat{L}_x, \hat{L}_y, \hat{L}_z]$ is the angular momentum operator, and the components of $\bsig = [\sig_x, \sig_y, \sig_z]$ are Pauli spin matrices. Angular states are labelled by the eigenvalue $\kappa$ which is related to the overall angular momentum $j$ by
\begin{equation}
\kappa = \pm(j + \halfb) = \left\{ \begin{array}{ll}  l  +  1  \quad \text{where } l \equiv j - 1/2 \\
                                                     -l  \quad  \text{where } l \equiv j + 1/2 
				\end{array} \right.    . 
\end{equation}
Note that our sign convention for $\kappa$ may differ from the literature \cite{Rose-1961}. Explicitly, the spherical spinors are
\begin{equation}
\chi_\kappa^\mu(\theta, \phi) = S_\kappa \sum_{m = \pm 1/2} C(l \halfa j; \mu - m, m) Y^{\mu - m}_l(\theta, \phi) \chi^m , 
\label{sphericalspinors}
\end{equation}
where $S_\kappa = \kappa / |\kappa|$, $C(j_1 j_2 j_{tot}; m_1 m_2)$ are the Clebsch-Gordan coefficients, $Y^m_l (\theta, \phi)$ are spherical harmonics, and $\chi^m$ are the spin-up and spin-down spinors, $\chi^{\halfb} = \left[ \begin{smallmatrix} 1 \\ 0 \end{smallmatrix} \right]$ and $\chi^{-\halfb} = \left[ \begin{smallmatrix} 0 \\ 1 \end{smallmatrix} \right]$.
%\begin{equation}
%\chi^{\halfb} = \begin{pmatrix} 1 \\ 0 \end{pmatrix}, \quad \quad \chi^{-\halfb} = \begin{pmatrix} 0 \\ 1 \end{pmatrix}
%\end{equation}
The positive and negative $\kappa$ spinors are related by
\begin{equation}
\sig_r \chi_\kappa^\mu = \chi_{-\kappa}^\mu , 
\end{equation}
where $\sigma_r = r^{-1} \sum_i x^i \sig_i $. 

%To simplify the analysis, we will henceforth assume that the momentum of the incident wave is aligned with the z-axis. Then
%\begin{equation}
%\chi^\mu_\kappa (\mathbf{e_3}) = \left( \frac{2l + 1}{4 \pi} \right)^{\halfb} C(l \halfa j; 0 \mu) \chi^\mu
%In Pauli theory, a plane wave solution incident along the z axis can be expressed as a sum of spherical waves, as
%\begin{equation}
%\chi^m e^{i p z} = (4 \pi)^\halfb \sum_{\kappa} i^l (2l + 1)^\halfb C(l \halfa j; 0, m) j_l(pr) \chi_\kappa^m
%\end{equation}
%where $j_l$ are spherical Bessel functions.

%Consider, for instance, the non-relativistic Hamiltonian for a spin-half particle in a central field, $H = \hat{p}^2 / 2m + V_1(r) + V_2(r) \mathbf{s} \cdot \hat{\mathbf{L}}$. The orbital angular momentum operators $L_i = \epsilon_{ijk} x_j \hat{p}_k$ do not commute with the Hamiltonian, whereas the operator $\hat{\kappa} = (\bsig \cdot \bL + 1)$.

%The operator $\hat{\kappa} = (\bsig \cdot \bL + 1)$, however, does commute with the Hamiltonian, so the eigenvalue $\kappa$ is a conserved quantum number. 

Consider, first, the free-particle Dirac equation, $i \hat{\gamma}^\mu
\partial_\mu \Psi - \mass \Psi = 0$. It admits plane
wave solutions; for example, a plane wave propagating in the z
direction can be written as
\begin{equation}
\Psi_\text{plane}^{\pm} = \begin{pmatrix} \chi^{\pm} \\
\frac{p}{E+m} \sigma_z \chi^{\pm} \end{pmatrix} e^{ipz}
e^{-iEt} .
\label{diracplanewave}
\end{equation}
It also admits separable solutions in spherical coordinates of the form
\begin{equation}
\Psi^\mu_\kappa = \frac{1}{r} \begin{pmatrix} u_1(r) \chi^\mu_\kappa \\ u_2(r) \chi^\mu_{-\kappa} \end{pmatrix} e^{-i E t} ,
\label{diracsphericalwave}
\end{equation}
where $u_1(r)$ and $u_2(r)$ are radial functions, and
$\chi^\mu_\kappa$ the Pauli two-component spherical spinors
(\ref{sphericalspinors}). The free particle radial equations can
be written in matrix form as
\begin{equation}
\frac{d}{dr} \begin{pmatrix} u_1 \\ u_2 \end{pmatrix} = 
\begin{pmatrix} \kappa / r & i(E + \mass) \\ i(E - \mass) & -\kappa / r \end{pmatrix}
\begin{pmatrix} u_1 \\ u_2 \end{pmatrix} , 
\end{equation}
and their solutions are spherical Bessel functions
\begin{equation}
u_1^{(\kappa)}(r) = r j_l (pr), \quad \quad u_2^{(\kappa)}(r) = \frac{i p S_\kappa}{E + m} r j_{\overline{l}}(pr) , 
\end{equation} 
where $\overline{l} = l + S_\kappa$. The plane wave solution (\ref{diracplanewave}) can be
written as a sum of spherical solutions (\ref{diracsphericalwave}), as
\begin{equation}
\Psi_{\text{plane}}^m =  e^{-iEt} (4 \pi)^{1/2} \sum_{\kappa \neq 0} i^l (2l + 1)^{1/2} C(l \halfa j; 0 m) \frac{1}{r} \begin{pmatrix} u_1^{(\kappa)} \chi^m_\kappa \\  u_2^{(\kappa)} \chi^m_{-\kappa} \end{pmatrix}. 
\end{equation}
The asymptotic behaviour of the radial solutions is found by noting
\begin{equation}
j_l(pr) \rightarrow \frac{1}{p r}\sin\left(pr - \frac{l \pi}{2}\right), \quad \quad \text{as } r \rightarrow \infty . 
\label{asymp-bessel}
\end{equation}

When an interaction is introduced, the unperturbed plane wave is no
longer an eigenstate of the Hamiltonian. Instead we look for a 
time-independent solution with the asymptotic behaviour
\begin{equation}
\Psi \rightarrow \Psi_\text{plane} + \frac{\bspin}{r} e^{ i p r }e^{-iEt} ,
\label{asymp-wavefn}
\end{equation}
where $\bspin$ is a four-component spinor to be determined. Asymptotically,
the wave function is interpreted as the sum of an incident plane wave
and a radially-outgoing scattered wave. The effect of an interaction
is to introduce phase shifts into the asymptotic solution
(\ref{asymp-bessel}). We define an interaction phase
shift $\delta_\kappa$ by comparing the asymptotic solution with the
free-space form (\ref{asymp-bessel}),
\begin{equation}
u_1^{(\kappa)}(r) \rightarrow p^{-1} \sin\left(p r - \frac{l \pi}{2} +
\delta_\kappa\right), \quad \text{as } r \rightarrow \infty . 
\end{equation}

An expansion of the form (\ref{asymp-wavefn}) is only strictly valid
if the interaction is localised. In a long-range potential, such as
the Coulomb $1/r$ potential, the asymptotic solutions are modified by
the presence of a radially-dependent phase factor. The same is true in
a gravitational field, as we show in the next section. The extra phase
factor means that incident plane waves are distorted by the presence
of the potential even at infinity. However, the extra phase shift is
independent of $\kappa$, so does not contribute to $\delta_\kappa$ and
the phase difference between partial waves.

Let us now assume that the wave is incident along the z-axis, and is
in a superposition of spin-up and spin-down states, with upper
components
\begin{equation}
\sum_{m=\pm \halfb} c_m \chi^m = \begin{pmatrix} c_{\halfb} \\ c_{-\halfb} \end{pmatrix} . 
\end{equation}
We now construct a partial wave series with asymptotic form
(\ref{asymp-wavefn}). For clarity we need only consider the upper
components of the scattered wave, given by
\begin{align}
\text{Upper} ( \bspin ) = \sum_{m \tau} c_m B_\tau^m \chi^\tau ,
\end{align}
where
\begin{equation}
B^m_\tau = \frac{(4\pi)^{1/2}}{2 i p} \sum_\kappa (2l + 1)^{1/2} (e^{2i \delta_\kappa} - 1) C(l \halfa j; 0 m) C(l \halfa j; m - \tau, \tau) Y^{m-\tau}_l (\bp_f) . 
\end{equation}
Using the properties of Clebsch-Gordan coefficients, it is
straightforward to show that $B^{\halfc}_{\halfc} =
B^{-\halfc}_{-\halfc} $. The scattered wave spinor can be written as
$A \sum_m c_m \chi^m$, where $A$ is a transition amplitude matrix,
with components
\begin{equation}
A = \begin{pmatrix} B^{\halfc}_{\halfc} & B^{-\halfc}_{\halfc} \\  B^{\halfc}_{-\halfc} & B^{-\halfc}_{-\halfc}  \end{pmatrix} = f(\theta) + i g(\theta) \bsig \cdot \mathbf{\hat{n}} .
\label{transitionamplitude}
\end{equation}
Here, $f$ and $g$ are complex amplitudes, and it can be shown (see,
for example, \cite{Rose-1961}) that the rotation vector
$\mathbf{\hat{n}}$ is orthogonal to both the incident and scattered
direction, $\mathbf{\hat{n}} = \bp_i \times \bp_f / |\bp_i \times
\bp_f|$. The scattered intensity from an unpolarized beam is the sum of
the squares of the amplitudes,
\begin{equation}
\frac{d \sigma}{d \Omega} = |f|^2 + |g|^2. 
\end{equation}
In terms of phase shifts, the scattering amplitudes are
\begin{align}
f(\theta) &= \frac{1}{2ip} \sum_{l = 0} \left[ (l+1)(e^{2i\delta_{l+1}} - 1)  + l (e^{2i\delta_{-l}} - 1) \right] P_l (x)
\label{scattering-amp-f} \\
g(\theta) &= \frac{1}{2ip} \sum_{l=0} (e^{2i\delta_{l+1}} - e^{2i\delta_{-l}}) P^1_l (x) .
\label{scattering-amp-g}
\end{align}
where $x = \cos \theta$. There are two limiting cases to consider. First, in the
non-relativistic limit we expect the interaction to
depend only on orbital angular momentum ($l$). In this case, we expect
$\delta_{l+1} \approx \delta_{-l}$, and the spin is then unchanged by the interaction, and
we recover the non-relativistic scattering series $(2ip)^{-1}
\sum_{l=0} (2l+1) e^{2i\eta_l} P_l (x)$ (where $\eta_l =
\delta_{l+1} = \delta_{-l}$).

Second, in the highly-relativistic regime $E \gg m$, the
phase shifts instead depend only on total angular momentum ($j$), and so
$\delta_{\kappa} \approx \delta_{-\kappa}$. Then, for $\theta \neq 0$,
\begin{align}
f(\theta) &\approx \frac{1}{2ip} \sum_{l = 0} (l+1) e^{2i\delta_{l+1}} \left[P_{l+1}(x) + P_l(x) \right] \label{fmassless}\\
g(\theta) &\approx \frac{1}{2ip} \sum_{l=0} e^{2i\delta_{l+1}} \left[ P_{l}^1(x) - P_{l+1}^1(x) \right] \label{gmassless}
\end{align}
Using the properties of the Legendre polynomials, it is straightforward to show (eg. Mott \cite{Mott-1965}) that 
\begin{equation}
g = \tan(\theta / 2) f 
\label{fgmassless}
\end{equation}
in the relativistic limit. Physically, this implies that the helicity
of a massless wave remains unchanged by the interaction.

If the particle has mass then the scattered beam may become partially
polarized by the interaction. The degree of polarization $\polP$
induced in an initially unpolarized beam is
\begin{equation}
\polP = -i \frac{f g^\ast - f^\ast g}{|f|^2 + |g|^2}  
\label{Mott-polarisation}
\end{equation}
in the direction $\hat{\mathbf{n}}$.  This effect is known as
Mott polarization, and is well known for the Coulomb interaction of an
electron with a nucleus. Conversely, if the incident beam is already
partially polarized (with polarization vector $\bP$) then the wave is
scattered asymmetrically, with asymmetry factor $\polP \bP \cdot
\hat{\mathbf{n}}$.
%A visual interpretation of the Mott polarisation effect is discussed
%in Appendix \ref{spinvector}.

\subsection{\label{sec:diracequation}The Dirac equation on a Schwarzschild background}
We now consider the Dirac equation on a Schwarzschild
spacetime. Before starting, we must choose an appropriate coordinate
system. A disadvantage of standard Schwarzschild coordinates is
that they are valid only in the exterior region, $r > 2M$. Geodesics
do not cross the horizon in a finite Schwarzschild coordinate time,
and all solutions to the wave equation are singular at the
horizon. Here, we prefer to work with coordinate systems that remove
the coordinate singularity at $r = 2M$. Two possibilities are advanced
Eddington-Finkelstein coordinates, with metric
\begin{equation}
d s^2 = (1 - 2M/r) d t^2 - (4M / r) dt \, dr  - (1 + 2M/r)  d r^2
 - r^2  d\Omega^2, 
\end{equation}
and Painlev\'e--Gullstrand coordinates~\cite{Doran-2002,Martel-2001}, with metric
\begin{equation}
d s^2 =  \left(1 - 2M/r \right) d t^2 - \sqrt{8M/r} dt \, dr  -  d r^2
- r^2 d\Omega^2.
\label{painleve}
\end{equation}
Both coordinate systems deal smoothly with the horizon, and cover
regions I and III of the Penrose diagram of the fully extended Kruskal
manifold~\cite{Birrell-1982}. For a discussion of gauge-invariance and
the choice of coordinates, see \cite{Lasenby-2005-bs}.  %These two
%regions are the ones of
%physical relevance for this paper.  

The time coordinate of the Painlev\'e-Gullstrand (PG) system has a
simple physical interpretation. It corresponds to the proper time as
measured by a observer in freefall who starts from rest at
infinity. For this reason, we now formulate the Dirac
equation on a black hole background described by PG coordinates. 

We let $\{\gam_0,\gam_1,\gam_2,\gam_3\}$ denote the gamma matrices in the
Dirac--Pauli representation, and introduce spherical polar coordinates
$(r,\theta,\phi)$.  From these we define the unit polar matrices
\begin{align}
\gam_r &=  \sin\!\theta (\cos\!\phi \, \gam_1 + \sin\!\phi \,
\gam_2) + \cos\!\theta\, \gam_3 \, , \nn \\
\gam_\theta &= \cos\!\theta (\cos\!\phi \, \gam_1 + \sin\!\phi \,
\gam_2) - \sin\!\theta\, \gam_3 \, ,\nn \\
\gam_\phi &= - \sin\!\phi \, \gam_1 + \cos\!\phi \,
\gam_2.
\end{align}
In terms of these we define four position-dependent matrices $\{g_t,
g_r, g_\theta, g_\phi \}$ by
\begin{align}
\qquad
g_t &= \gam_0 + \sqrt{\frac{2M}{r}} \gam_r \, , &
g_\theta &= r \gam_\theta \, , \nn \\
g_r &= \gam_r \, , &
g_\phi &= r \sin\!\theta \gam_{\phi} .
\qquad
\end{align}
These satisfy the anti-commutation relations
\begin{equation}
\{ g_\mu, g_\nu \}_+ = 2 g_{\mu\nu} I 
\end{equation}
where $g_{\mu\nu}$ is the Painlev\'e--Gullstrand metric of
equation~(\ref{painleve}).  The reciprocal matrices $g^t, g^r,
g^\theta, g^\phi $ are defined by the equation
\begin{equation}
\{ g^\mu, g_\nu \}_+ = 2 \delta^\mu_\nu I,
\end{equation}
and both sets are well-defined everywhere except at the origin. 

The Dirac equation for a spin-half particle of mass $\mass$ is
\begin{equation}
i g^\mu \grad_\mu \psi = \mass \psi,
\end{equation}
where 
\begin{equation}
\grad_\mu \psi = (\partial_\mu + \frac{i}{2} \Gamma^{\alpha \beta}_{\mu}
\Sigma_{\alpha \beta}) \psi,
\qquad
\Sigma_{\alpha \beta} = \frac{i}{4}[\gamma_\alpha, \gamma_\beta]_-.
\end{equation}
The components of the spin connection are found in the standard
way~\cite{Nakahara-1990} and are particularly simple in the
Painlev\'e--Gullstrand gauge~\cite{Lasenby-2005-bs},
\begin{equation}
g^\mu  \frac{i}{2} \Gamma^{\alpha \beta}_{\mu}
\Sigma_{\alpha \beta} = -\frac{3}{4r} \sqrt{\frac{2M}{r}} \gam_0.
\end{equation}
An advantage of our choice of metric is that the Dirac equation can
now be written in a manifestly Hamiltonian form
\begin{equation}
i \!\! \delsl \psi - i \gamma^0 \left(\frac{2M}{r} \right)^{1/2}
\left( \frac{\partial}{\partial r}  + \frac{3}{4r} \right) \psi = m \psi,
\label{dirac-hamiltonian}
\end{equation}
where ${\not\!\partial}$ is the Dirac operator in flat Minkowski
spacetime. The interaction term is non-Hermitian, as the singularity
acts as a sink for probability density, making absorption possible.

The Dirac equation is separable in time, with solutions that
go as $\exp(-iEt)$. The energy $E$ conjugate to time-translation is
independent of the chosen coordinate system, and has a physical
definition in terms of the Killing time~\cite{Lasenby-2005-bs}. We 
exploit the spherical symmetry to separate the spinor into
\begin{equation}
\psi = \frac{e^{-iEt}}{r} 
\begin{pmatrix}
u_1(r) \chi_\kappa^\mu (\theta, \phi) \\
u_2(r) \chi_{-\kappa}^\mu (\theta, \phi) 
\end{pmatrix}
\label{trispn}
\end{equation}
The trial function (\ref{trispn}) results in a pair of coupled
first-order equations  
\begin{multline}
\left(1 - 2M / r \right) \frac{\ud}{\ud r} 
\begin{pmatrix} u_1 \\ u_2 \end{pmatrix} = 
\begin{pmatrix} 1 & \sqrt{2M / r} \\ \sqrt{2M / r} & 1 \end{pmatrix}
 \\ 
\cdot
\begin{pmatrix} 
\kappa / r  &  i (E+m) - (2M/r)^{1/2}(4r)^{-1} \\  
i (E-m) - (2M/r)^{1/2}(4r)^{-1} &  -\kappa / r 
\end{pmatrix}
\begin{pmatrix} u_1 \\ u_2 \end{pmatrix}.
\label{matrixeqn}
\end{multline}
The equations have regular singular points at the origin and horizon,
and an irregular singular point at $r = \infty$. Analytic solutions to
radial equations of this nature have been investigated by Leaver
\cite{Leaver-1986} and others \cite{Mano-1996}. Here, we prefer to use
series solutions around the singular points as initial data for a
numerical integration scheme.

\subsection{Series Solutions and Boundary Conditions}
As is clear from (\ref{matrixeqn}), there is a regular singular point
in the coupled equations at the horizon, $r = 2M$. We look for series
solutions
\begin{equation}
U_\text{hor} =
\begin{pmatrix} u_1 \\ u_2 \end{pmatrix} 
= (r-2M)^s \sum_{n=0}^{\infty} \begin{pmatrix} a_n \\ b_n \end{pmatrix}
(r-2M)^n  
\end{equation}
where $s$ is the lowest power in the series, and $a_n$, $b_n$ are
coefficients to be determined. On substituting into (\ref{matrixeqn})
and setting $r = 2M$ we obtain an eigenvalue equation for $s$, which
has solutions
\begin{equation}
s = 0 \hspace{0.5cm}\text{ and }\hspace{0.5cm}s = -\half + 4 i M E. 
\end{equation}
The regular root $s=0$ ensures that we can construct solutions which
are finite and continuous at the horizon. Regular solutions
automatically have an ingoing current at the horizon. The singular
branch gives rise to discontinuous, unnormalisable solutions with an
outgoing current at the horizon \cite{Lasenby-1998-gtg}.  We therefore
restrict attention to the regular, physically-admissable
solutions. The eigenvector for the regular solution is
\begin{equation}
\begin{pmatrix} a_0 \\ b_0 \end{pmatrix} = 
\begin{pmatrix} \kappa - 2 i M (E+m) + 1/4  \\ \kappa + 2 i M (E-m) - 1/4 
\end{pmatrix} .
\end{equation}

In order to expand about infinity we need to take care of the
irregular singularity present there. There are two sets of solutions,
$U^\text{(out)}_\infty$ and $U^{\text{(in)}}_\infty$, which asymptotically resemble
outgoing and ingoing radial waves with additional radially-dependent
phase factors. To lowest order,
\begin{align}
U^\text{(out)}_\infty &= 
e^{i p r} e^{i (\phi_1 + \phi_2)} 
\begin{pmatrix} 1 \\ p / (E+m) \end{pmatrix} 
\nn \\
U^\text{(in)}_\infty &= 
e^{-i p r} e^{i (\phi_1 - \phi_2)} 
\begin{pmatrix} 1 \\ -p / (E+m) \end{pmatrix} 
\label{asympt1}
\end{align}
where the phase factors $\phi_1(r), \phi_2(r)$ are given by
\begin{equation}
\phi_1(r) = E \sqrt{8Mr},  \qquad  
\phi_2(r) = \frac{M}{p} \left(m^2 + 2p^2 \right) \ln(pr) .
\end{equation}
The general solution as $r \rightarrow \infty$ is
a superposition of the ingoing and outgoing waves,
\begin{equation}
U = \left\{ \begin{array}{ll}
  U^{\text{(in)}}_\text{hor} & r \rightarrow 2M \\
  \Ainalt U^{\text{(in)}}_\infty + \Aoutalt U^{\text{(out)}}_\infty  & r \rightarrow \infty
\end{array}\right.
\label{asympt2}
\end{equation}
%\begin{align}
%U(r \rightarrow 2M) &= U^{\text{(in)}}_\text{hor} \\
%U(r \rightarrow \infty) &= \Ainalt U^{\text{(in)}}_\infty + \Aoutalt U^{\text{(out)}}_\infty 
%\label{asympt2}
%\end{align}
%
for each angular mode. The magnitudes of $\Ainalt$ and
$\Aoutalt$ determine the amount of scattered and absorbed
radiation present. The phase shift $\delta_\kappa$ is given by their ratio,
\begin{equation}
e^{2i\delta_\kappa} = (-1)^{l+1} \frac{\Aoutalt}{\Ainalt}.
\label{phaseshiftdef}
\end{equation}

\subsection{Numerical Method}
To calculate the scattering amplitude we must first determine the
phase shifts. Various analytical methods for calculating phase shifts
have been successfully applied to the black hole scattering problem, for waves
of spin $s = 0$ and $s = 2$. \cite{Matzner-1985, Anninos-1992,
Andersson-1994, Andersson-1995}. Here we calculate the phase shifts
using a numerical method. It is a relatively straightforward task to
integrate the radial equation (\ref{matrixeqn}) and apply the ingoing
boundary conditions (\ref{asympt2}) to compute the phase shifts
defined by (\ref{phaseshiftdef}). We determine the ingoing and
outgoing coefficients $\Ainalt$ and $\Aoutalt$ by matching the horizon
solution onto the infinity solution. In practice, this involves
choosing two boundary points $\rmin$ and $\rmax$, close to the horizon
and infinity respectively. We start at $\rmin$ with series solution
$U_\text{hor}(\rmin)$, and integrate out to $\rmax$ where we compare
the numerical solution with the asymptotic form $\Ainalt
U_\infty^\text{(in)}(\rmax) + \Aoutalt
U_\infty^\text{(out)}(\rmax)$. The phase shift calculated by this
method is independent of our choice of $\rmin$ and $\rmax$, provided
that they are sufficiently close to the boundary points. The phase
shift should also be independent of the choice of coordinates. We
repeated our analysis using Eddington-Finkelstein coordinates to check
the numerical accuracy of our results.

For the scalar wave in the large-$l$ limit \cite{Futterman-1988} the
phase shifts approach ``Newtonian'' values which arise in a strict
$1/r$ potential,
\begin{equation}
e^{2 i \delta_l^N} = \frac{\Gamma(l + 1 - i\beta)}{\Gamma(l + 1 + i \beta)},
\end{equation}
where $\beta = M (2 E^2 - \mass^2) / p = ME (1+v^2)/v$. Phase shifts of this form
also occur in the non-relativistic Coulomb scattering problem (see,
for instance, \cite{Mott-1965}), with $\beta_C = Z \alpha \mass /
p$. It is well known that the problem of non-relativistic Coulomb
scattering can be solved exactly in parabolic coordinates (see, for
example, \cite{Wu-1962}). The pure-Newtonian phase
shift series defined by
\begin{equation}
f_N(\theta) = \frac{1}{2 i p} \sum_{l=0} (2l + 1) \left[\frac{\Gamma(l + 1 - i\beta)}{\Gamma(l + 1 + i \beta)} - 1 \right] P_l( \cos \theta )
\label{newtonianseries}
\end{equation}
can be summed using the Coulomb result to
\begin{equation}
f_N(\theta) = \frac{\beta}{2 p} \frac{\Gamma(1 - i \beta)}{\Gamma(1 + i \beta)} \left[ \sin(\theta / 2) \right]^{-2 + 2 i \beta}.
\end{equation}
It is more difficult to sum the series (\ref{newtonianseries})
\emph{directly}, because it is poorly convergent. This is related to the
fact that an infinite number of Legendre polynomials are required to
describe the divergence at $\theta = 0$. Many years ago, Yennie,
Ravenhall and Wilson \cite{Yennie-1954} outlined one possible way
around this problem. Given a Legendre polynomial series
\begin{equation}
f(\theta) = \sum_{l=0} a_l^{(0)} P_l (\cos \theta)
\end{equation}
that is divergent at $\theta = 0$, one may define the $m$th reduced series,
\begin{equation}
(1 - \cos\theta)^m f(\theta) = \sum_{l=0} a_l^{(m)} P_l(\cos \theta).
\end{equation}
The reduced series is obviously less divergent at $\theta = 0$, so one
may hope that the reduced series converges more quickly. Using the
properties of the Legendre polynomials, it is straightforward to show
that the new coefficients $a_l^{i+1}$ are related to the old
coefficients $a_l^{i}$ by the iterative formula
\begin{equation}
a_l^{(i+1)} = a_l^{(i)} - \frac{l+1}{2l+3} a_{l+1}^{(i)} - \frac{l}{2l-1} a_{l-1}^{(i)}.
\end{equation}
We have found that this to be an excellent method for summing the series
numerically, and that two or three iterations are sufficient.

\section{\label{sec:results}Numerical Results}
In this section we present the results of our numerical calculations,
and compare with theoretical expectations outlined in sections
\ref{sec:classical} and \ref{sec:analytic}.

\subsection{Absorption Cross Sections}
The existence of an ingoing current at the horizon implies that flux is absorbed by
the black hole. The absorption cross section $\sigabs$ is defined as ratio of the
absorbed and incident fluxes. For the spin-half wave, 
\begin{equation}
\sigabs = \frac{\pi}{p^2} \sum_{\kappa \neq 0} |\kappa| \left(1 - |e^{2i\delta_\kappa}|^2 \right).
\end{equation}
At low energies, absorption is dominated by the $j = s$ mode. For
massless waves, the low energy limits are: (i) $\sigabs = 16 \pi
M^2$ for $s = 0$, (ii) $\sigabs = 2 \pi M^2$ for $s = 1/2$, (iii)
$\sigabs = \tfrac{64}{3}\pi(ME)^2 M^2$ for $s = 1$, and (iv) $\sigabs = \tfrac{256}{45}\pi (ME)^4 M^2$ for  $s = 2$ (see \cite{Starobinskii-1973}). Note that the low-energy scalar and spinor cross sections are proportional to the black hole area, whereas the electromagnetic and gravitational cross sections tend to zero in this limit.

\begin{figure}
\begin{center}
\includegraphics[width=8cm]{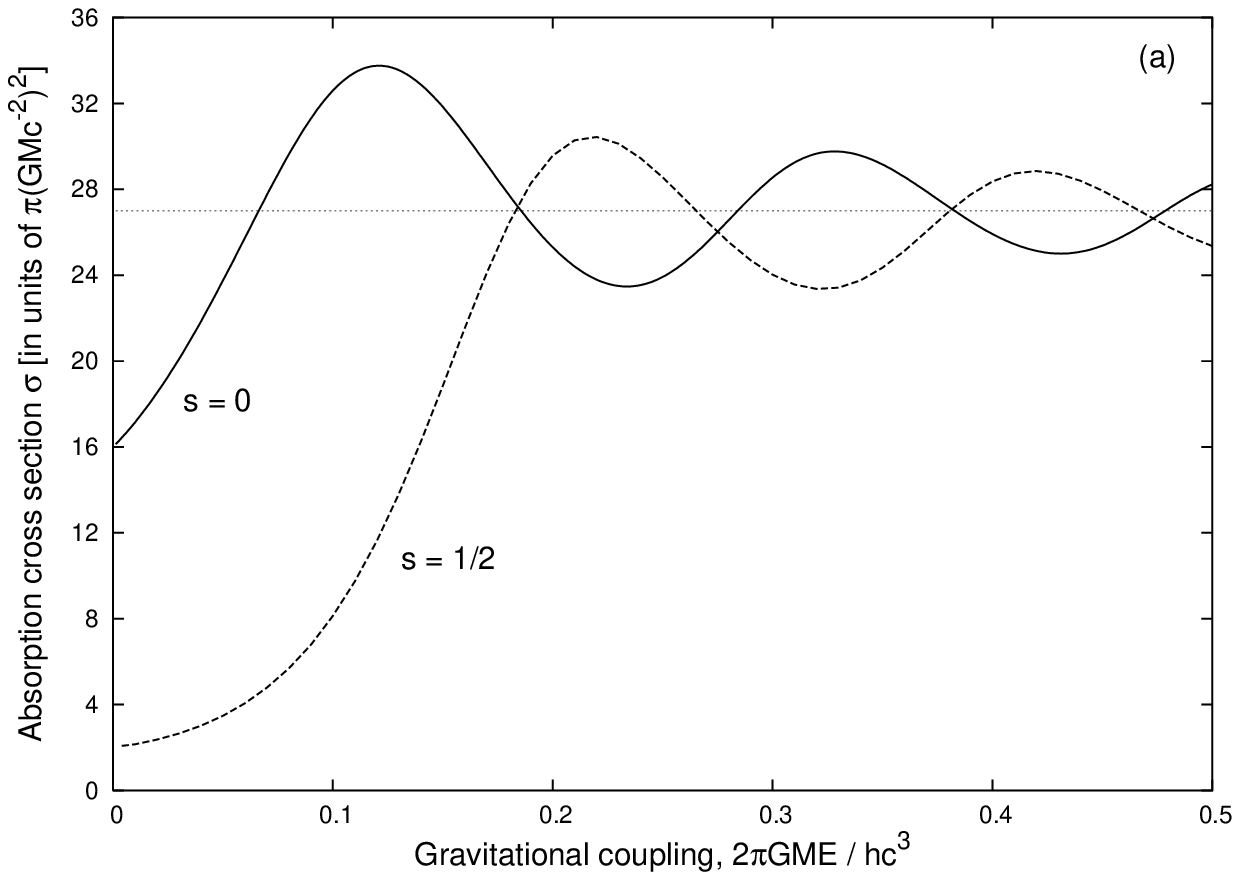}
\includegraphics[width=8cm]{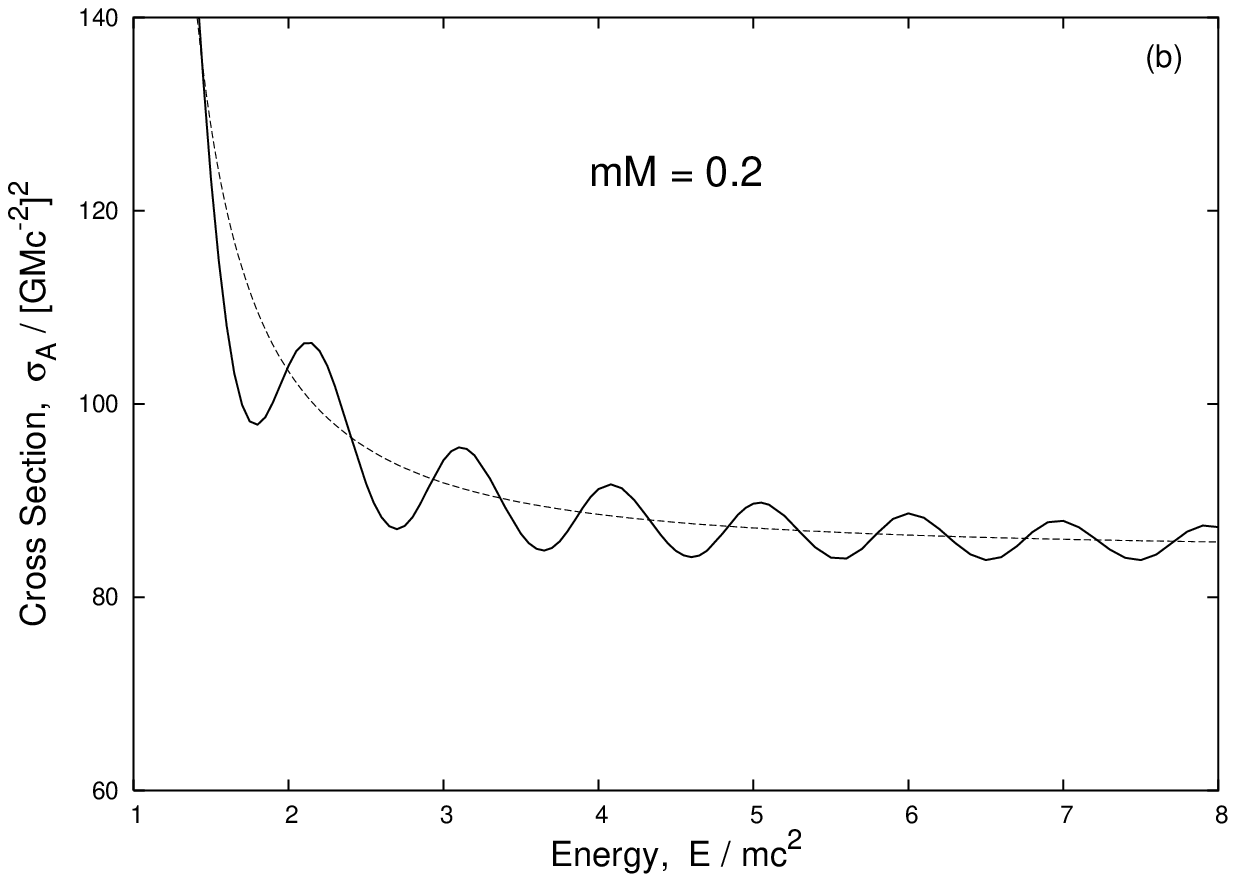}
\end{center}
\caption[dummy1]{Absorption cross section. (a) \emph{Massless}. Shows the absorption cross section of a massless scalar [solid] and a massless fermion [dashed]. The classical limit at $27 \pi$ is also marked.  (b) \emph{Massive}. The solid line shows the quantum absorption cross section for $GmM / \hbar c = 0.2$, and the dotted line is the classical cross section (\ref{classicalbcrit}).}
\label{absorption}
\end{figure}

The massless scalar and fermion cross sections are shown in the left panel of Fig
\ref{absorption}. For massive particles, the cross section diverges in
the low energy limit, as shown in the right panel of Fig
\ref{absorption}. In all cases, the cross section approaches the
geometrical optics value of $\sigabs = 27 \pi M^2$ in the high-energy
limit.  For more details on the absorption cross sections of massive
spin-half particles, see Doran \emph{et al}.  \cite{Doran-2005-abs}.

\subsection{Scattering at low couplings, $\lambda \gg r_s$}
\begin{figure}
\begin{center}
\includegraphics[width=8.6cm]{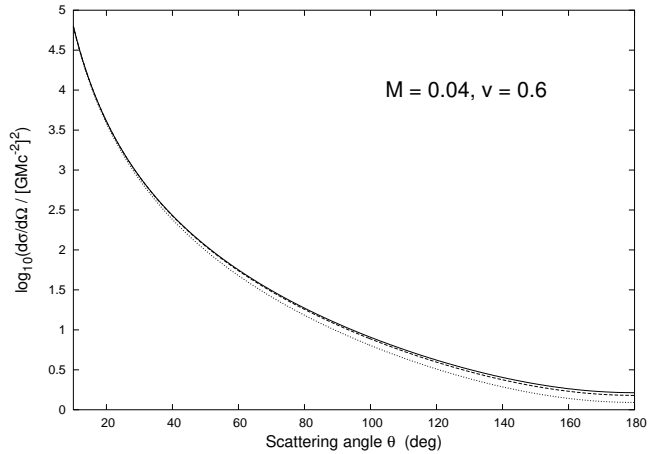}
\end{center}
\caption[dummy1]{Comparison of the numerical cross section [solid]
with the first-order [dotted] and second-order [dashed] Born
approximation at $ME = 0.04$, $v = 0.6$. Note that the scale on the y-axis of the
left plot is logarithmic.}
\label{born-approx}
\end{figure}

In section \ref{sec:analytic} we outlined our expectations for
scattering at low couplings. We now test them
numerically. Figure \ref{born-approx} compares the perturbation series
(\ref{born-series}) with a numerically-calculated cross section, at
$ME = 0.04$ and $v = 0.6$. It is clear that the approximation is
improved by the inclusion of the second-order term,
Eq. (\ref{second-order-scat}).

\begin{figure}
\begin{center}
\includegraphics[width=16cm]{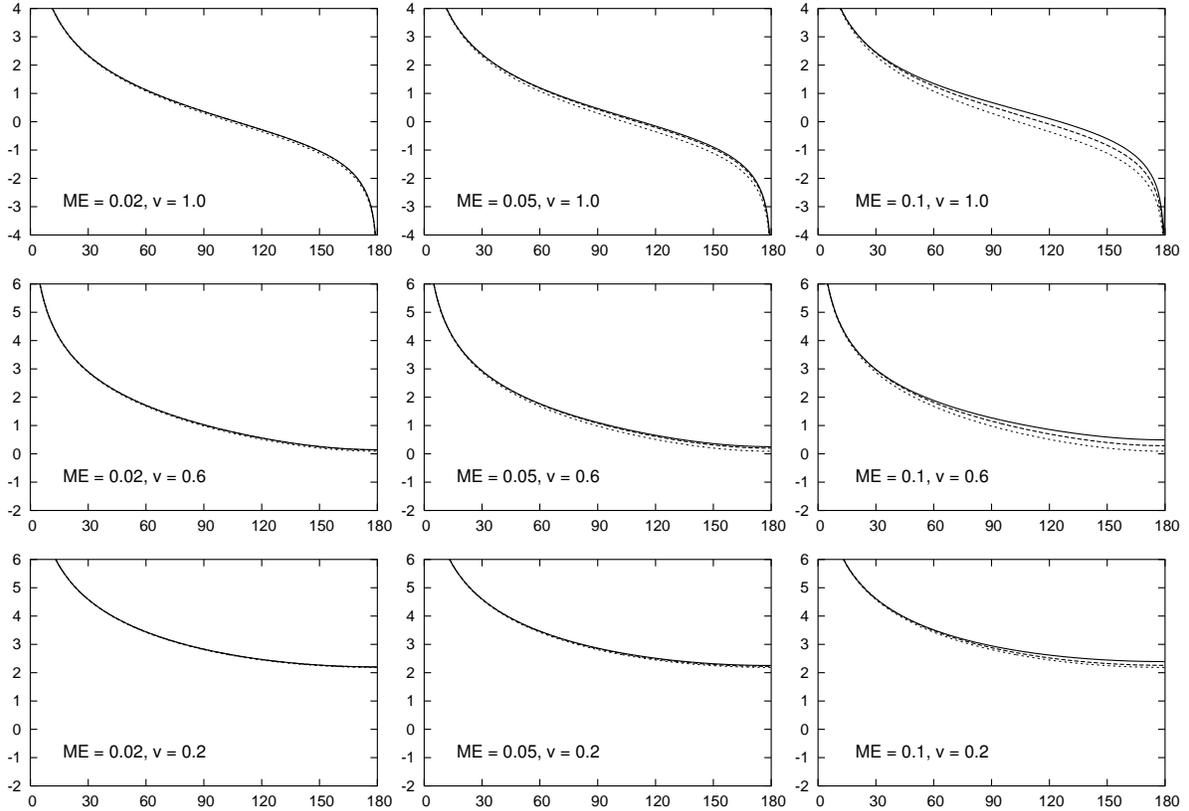}
\end{center}
\caption[dummy1]{Comparison of the numerical cross section [solid]
with the first-order [dotted] and second-order [dashed] Born
approximation across a range of $ME$ and $v$. The $x$-axis shows the
scattering angle (deg) and the $y$-axis shows the cross section,
$\log_{10} \left(M^{-2} \frac{d \sig}{d \Omega} \right)$.}
\label{born-approx-2}
\end{figure}

Figure \ref{born-approx-2} compares the perturbation results with the
numerical calculations across a range of values of $ME$ and $v$. The
first-order series (\ref{first-order-scat}) is shown as a dotted line,
and the second-order series (\ref{second-order-scat}) is shown as a
dashed line. As expected, we find that the second-order correction
improves the fit, and the approximation is in excellent agreement when
$MEv \ll 0.1$.
At higher couplings, the truncated perturbation series is still accurate for
small scattering angles, but it fails to predict 
glory and spiral scattering oscillations. In this regime, $r_s \sim \lambda$, a full
numerical calculation is necessary.

\begin{figure}
\begin{center}
\includegraphics[width=8cm]{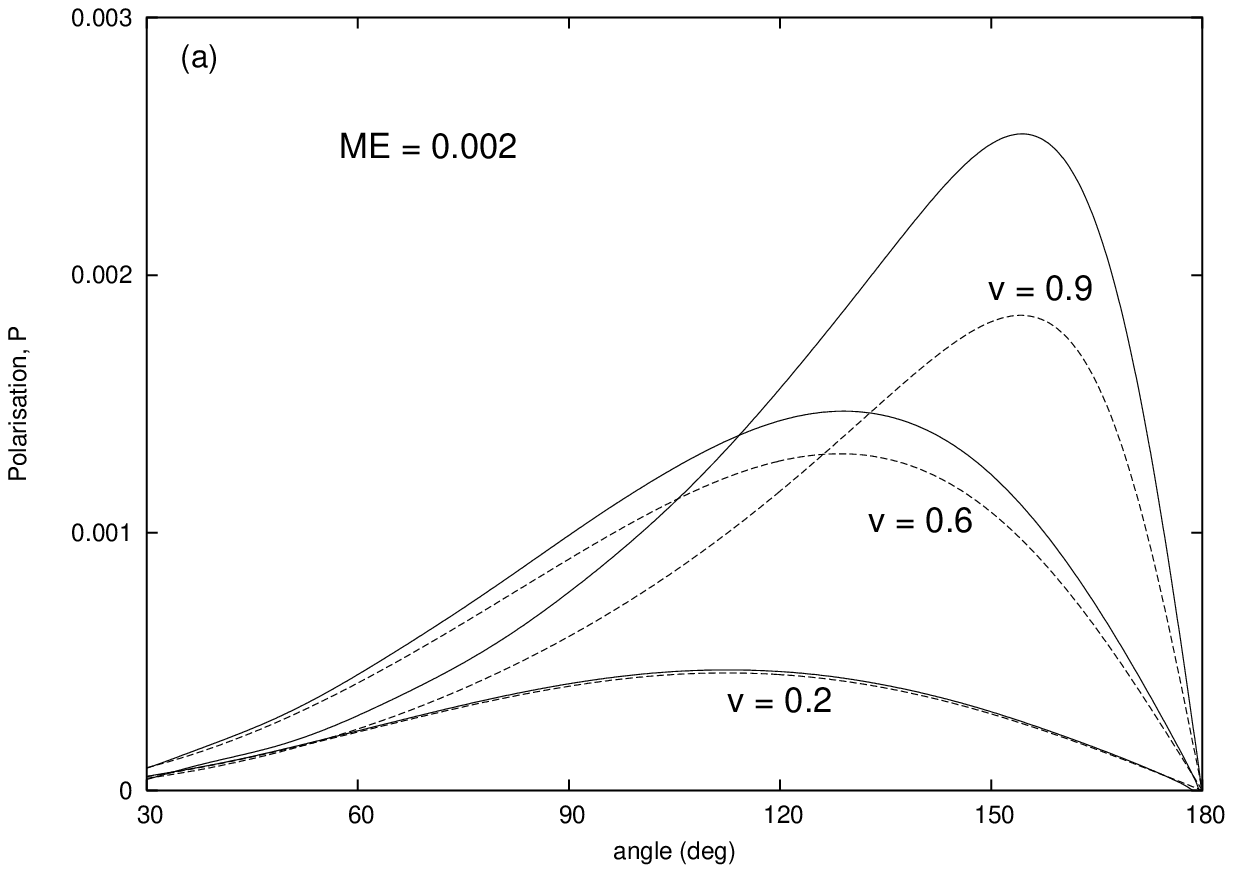}
\includegraphics[width=8cm]{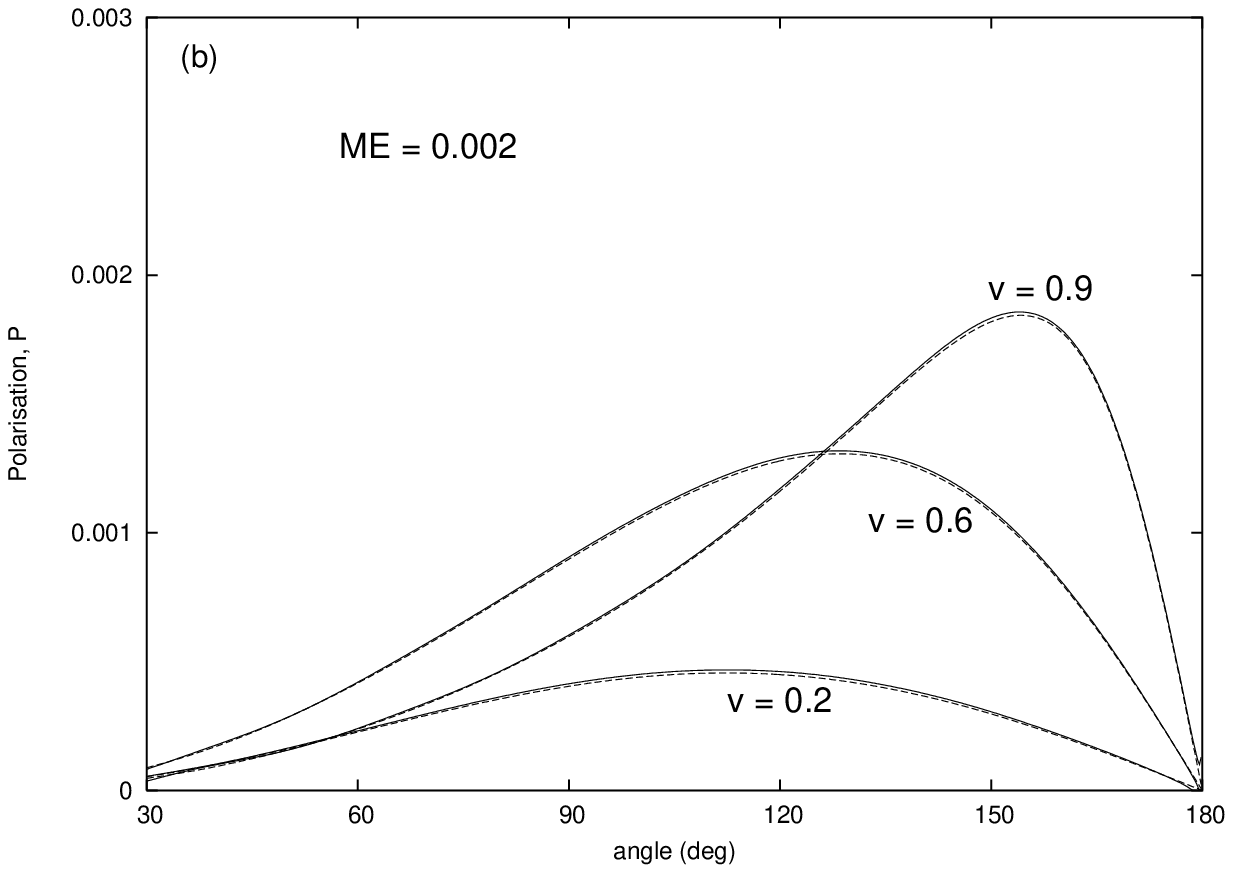}
\end{center}
\caption[dummy1]{Comparison of the numerical polarization [solid] and
the perturbation calculation (\ref{second-order-pol}) [dotted] for $ME = 0.002$. (a) With absorption. The perturbation result (\ref{second-order-pol}) does not match the numerical result. (b) Without absorption. Enforcing the artifical condition that numerical phase shifts are purely real removes the effect of absorption (see text). We then observe an excellent agreement with the perturbation result.}
\label{born-polarisation-fig}
\end{figure}

The perturbation calculation also predicts the magnitude of Mott
polarization, which we now test against numerical results. The
first-order polarization is given by (\ref{second-order-pol}), whereas
the numerical polarization is determined by
(\ref{Mott-polarisation}). Figure \ref{born-polarisation-fig} compares
the perturbation series and numerical results for a low coupling, $ME
= 0.002$. The left-hand plot (a) shows that there is an order-of-magnitude
agreement between the two results. Na\"ively, in the limit $ME
\rightarrow 0$, we would expect to find perfect agreement between the
two. We do not observe this. However, it is remarkable that we
\emph{can} obtain excellent agreement by artificially removing the
effects of absorption from the numerical results. If we set the
imaginary part of the numerical phase shifts to zero, we obtain the
right-hand plot, Fig. \ref{born-polarisation-fig}(b). It is 
apparent therefore that the perturbation method is somewhat flawed, as it
does not account for absorptive effects. This is related to the
fact that, whereas plane waves form a complete basis for the hydrogen
atom, they do not form a complete basis on a black-hole spacetime. 
We discuss this problem further in Section \ref{sec:discussion}.

\begin{figure}
\begin{center}
\includegraphics[width=8.6cm]{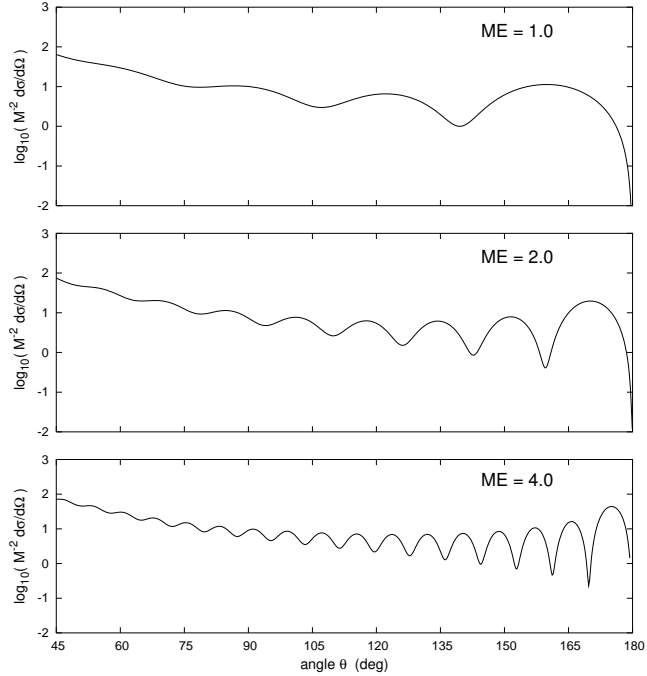}
\end{center}
\caption[dummy1]{Glory scattering cross sections for the massless spin-half wave for various couplings $ME \equiv GME / \hbar c^3$. }
\label{spinor-glory-1}
\end{figure}

\subsection{Scattering at higher couplings, $\lambda \sim r_s$}
\begin{figure}
\begin{center}
\includegraphics[width=8.6cm]{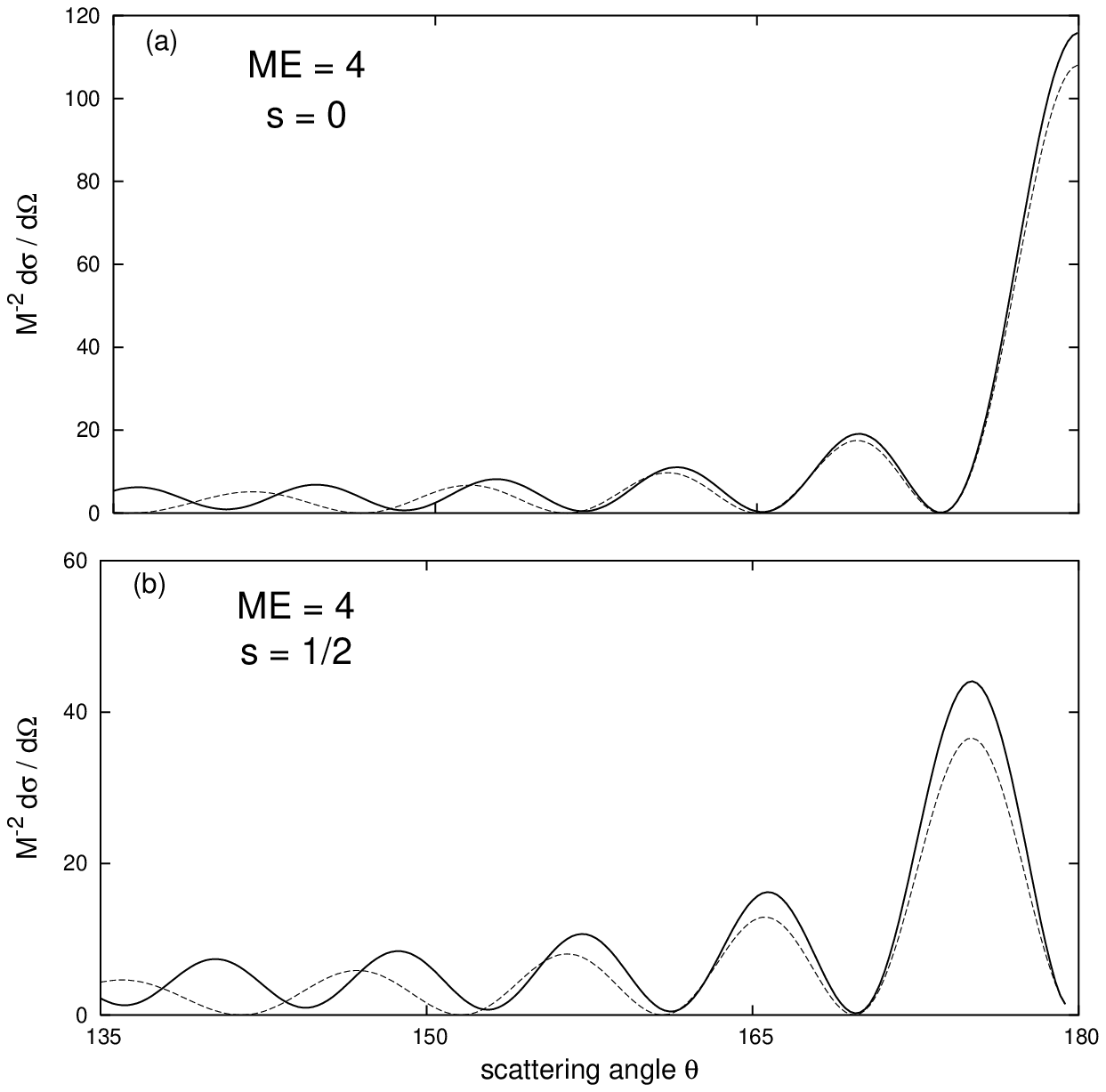}
\end{center}
\caption[dummy1]{Glory oscillations for the massless spinor and scalar waves. This plot compares the numerical glory oscillations (solid) with the approximate formula in the text (\ref{glory-approximation}) (dotted). (a) Scalar, $s = 0$. (b) Spinor, $s = 1/2$.}
\label{glory-approx}
\end{figure}

Figure \ref{spinor-glory-1} shows numerically-determined scattering
cross sections for a massless fermion at larger couplings. As
expected, we observe spiral scattering oscillations, and a zero
on-axis in the backward direction. The magnitude of the oscillations,
and their angular frequency, increases with black hole mass. The zero
in the backward direction can be verified by substituting $P_l(\pi) =
(-1)^l$ into the partial wave series (\ref{fmassless}). The massless
glory approximation (\ref{glory-approximation}) of Zhang and
DeWitt-Morette \cite{Zhang-1984} proves a good fit to the numerical
results close to $\theta = 180^\circ$, as shown in Fig.
\ref{glory-approx}. It correctly predicts the approximate magnitude
and angular width of the oscillations.

\begin{figure}
\begin{center}
\includegraphics[width=8.6cm]{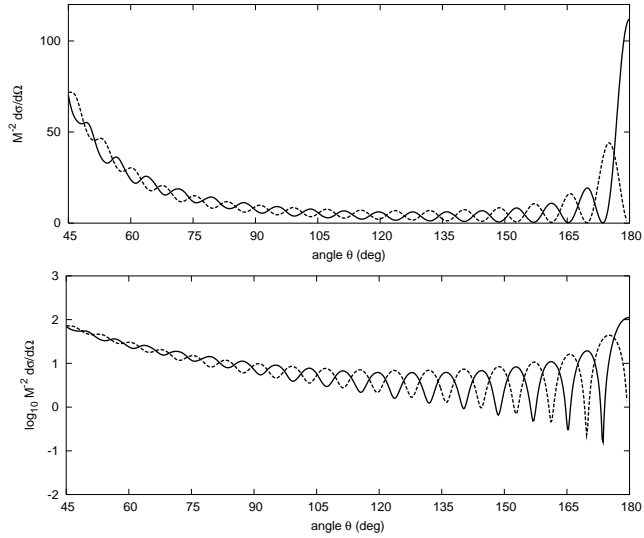}
\end{center}
\caption[dummy1]{Scattering cross sections for the massless scalar [solid] and spinor [dashed] waves at a coupling of $ME = 4$. Note that the bottom plot shows the same data as the top plot but with a logarithmic scale on the y-axis.}
\label{scalar-spinor-glory}
\end{figure}

The scattering cross sections for massless scalar and spinor waves are compared in Fig. \ref{scalar-spinor-glory}. In the backward direction the scalar wave is at a maximum on-axis, whereas the spinor wave is at a minimum, as predicted by the glory approximation. At intermediate angles, the regular oscillations in the intensity of the scalar and spinor waves appear to be $180^\circ$ out of phase. 

\begin{figure}
\begin{center}
\includegraphics[width=8.6cm]{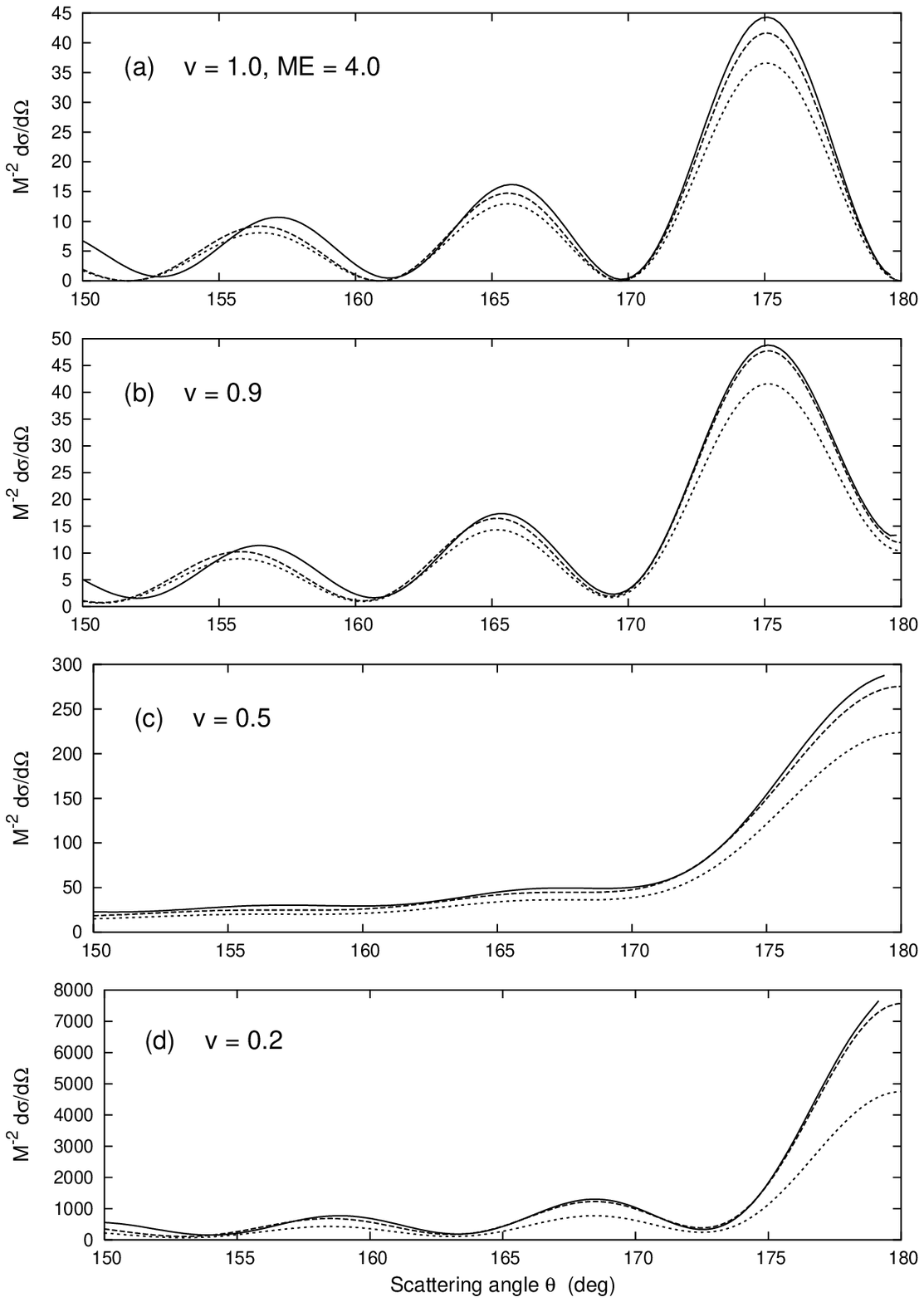}
\end{center}
\caption[dummy1]{Glory oscillations for the massive spin-half wave, over a range of particle speeds $v$.  The solid line shows the numerical cross section. The dashed line shows the result of the semi-classical approximation (\ref{glory-approx-massive-eq}), and the dotted line shows the result of using the modified Darwin approximation to estimate the glory impact parameter.}
\label{glory-approx-massive}
\end{figure}

Scattering cross sections for the massive spin-half wave are shown in
Fig. \ref{glory-approx-massive}. Here, the solid line shows the
partial-wave result, and the broken lines show the semi-classical
approximations to the glory (\ref{glory-approx-massive-eq}). The dotted
line is the result of using the modified Darwin approximation (\ref{darwin2}),
and the dashed line is calculated by integrating the orbit equations
numerically. It is clear that the numerical approach (dashed) provides
the better fit. This is because, for slower particles, we find $b_{glory}$ is too
far from $b_{c}$ for the first-order Darwin approximation to be
sufficient. In general, the Darwin approximation underestimates the
magnitude of the glory.

\begin{figure}
\begin{center}
\includegraphics[width=8cm]{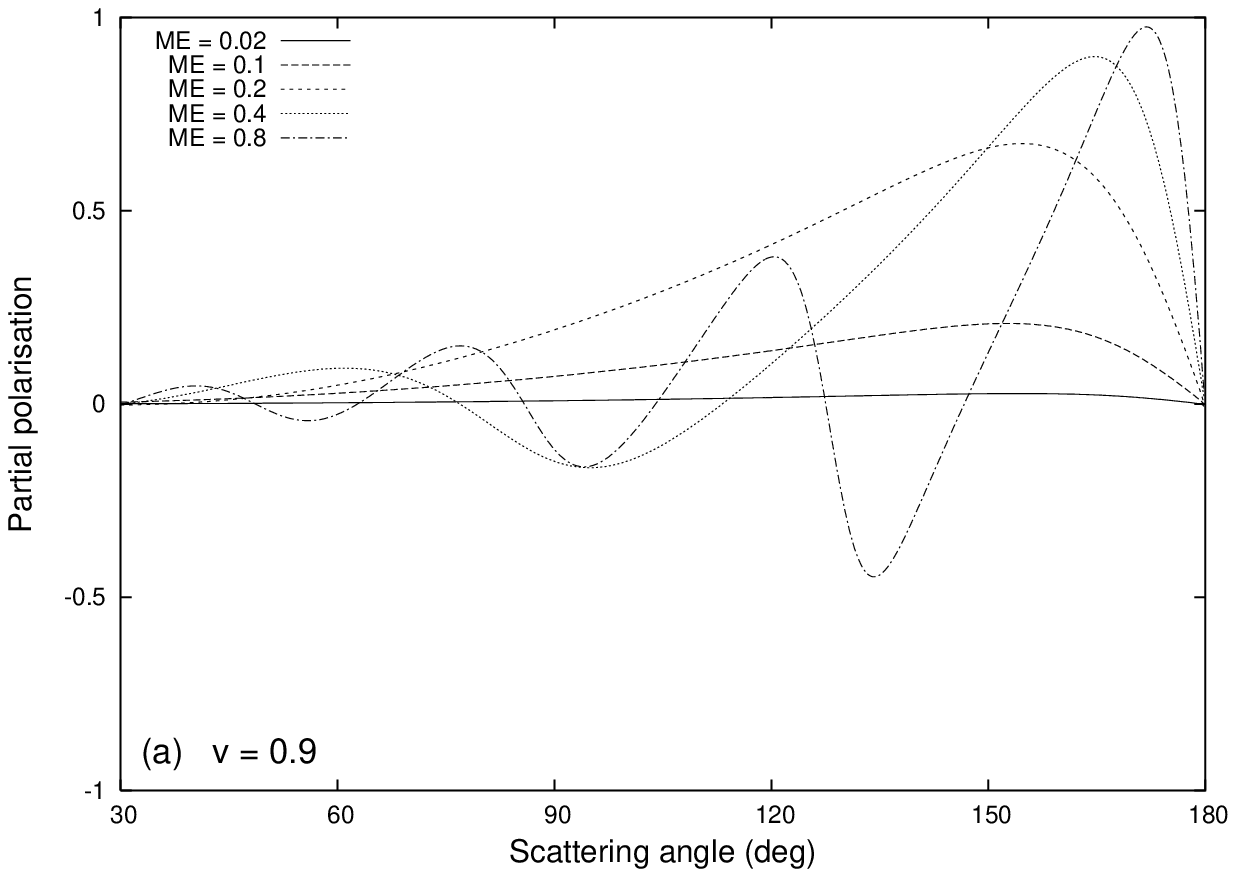}
\includegraphics[width=8cm]{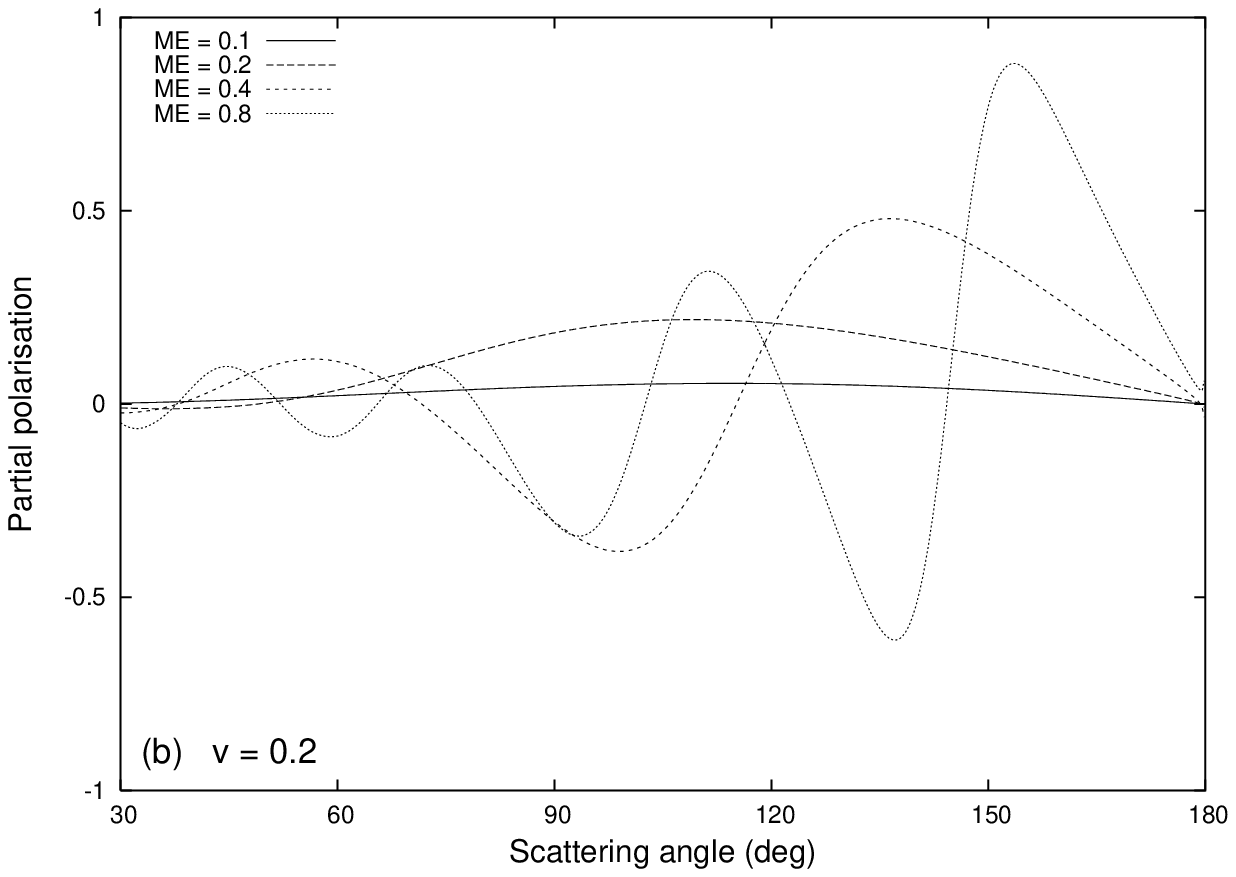}
\end{center}
\caption[dummy1]{Mott polarization as a function of scattering angle for a range of $ME$. 
Waves with mass are partially polarized by the interaction. As the coupling $ME$ is increased, oscillations arise in the polarization. (a) For $v = 0.9$. (b) For $v = 0.2$.}
\label{polarisation-angles}
\end{figure}

\begin{figure}
\begin{center}
\includegraphics[width=8.6cm]{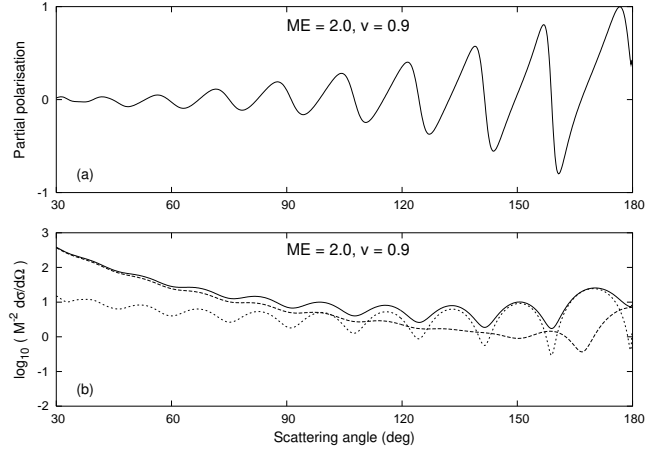}
\end{center}
\caption[dummy1]{Glory oscillations and polarization for $ME = 2$ and $v = 0.9$. (a) Partial polarization, $P$. (b) Scattering cross section. The solid line shows the total cross section, and the dotted and dashed lines show the contributions from the amplitudes $f$ and $g$ defined by Eq. (\ref{scattering-amp-f}) and (\ref{scattering-amp-g}).}
\label{polarisation-multiplot}
\end{figure}

Figures \ref{polarisation-angles}(a) and \ref{polarisation-angles}(b)
show the magnitude of Mott polarization as a function of scattering
angle, for a range of couplings. The relationship between
polarization, scattering angle, particle velocity, and black hole mass
is not immediately obvious. However, Fig. \ref{polarisation-multiplot}
shows that the oscillations in polarization are clearly related to the
glory and spiral scattering oscillations,
suggesting a semiclassical model for Mott polarization may be
possible. This awaits further work.

\section{\label{sec:discussion}Discussion and Conclusions}
In this paper we have presented an investigation into the scattering
and absorption of massive spin-half waves by a Schwarzschild black
hole. We extended the approximations of Darwin \cite{Darwin-1959} and
Zhang and DeWitt-Morette \cite{Zhang-1984} to derive a semiclassical ($\lambda \ll r_s$)
approximation for massive fermion glory scattering. We also took the
perturbation analysis of Doran and Lasenby \cite{Doran-2002} to next
order, to derive new formulae for the scattering cross section and
Mott polarization in the low coupling limit, $\lambda \gg
r_s$. Finally, we applied partial-wave techniques to solve
the Dirac equation on the Schwarschild spacetime, and presented our
numerical results. We showed that our approximations provide good fits
to the numerical results in the appropriate limits.

We showed that non-zero particle mass leads to a phenomenon familiar
from electromagnetic scattering: Mott polarization. This is a
partial polarization created in the direction orthogonal to the
scattering plane. In the electromagnetic case, an estimate of the polarization
arises from the second-order term in the Born approximation
\cite{Dalitz-1951}. In the gravitational case, a similar estimate
arises from our second-order perturbation analysis in the Kerr-Schild
gauge, Eq. (\ref{second-order-pol}). However, our estimate only agrees with
the numerical calculation if we artificially suppress black hole
absorption (see Fig. \ref{born-polarisation-fig}). This highlights a
shortcoming in the Born approximation method: it cannot account for
absorption effects. To see why, consider the Hamiltonian form of the
Dirac equation in PG coordinates, Eq.  (\ref{dirac-hamiltonian}). As
$r \rightarrow 0$, all solutions go as $\psi \sim r^{-3/4}$, and have
a net ingoing current. Clearly, such solutions cannot be described by
a linear sum of plane waves used in the Born approximation. A related
observation is that the Hamiltonian is not Hermitian as probability
flux is removed at the origin \cite{Lasenby-2005-bs}. It is not
immediately clear how we extend the Born approximation to such open
systems. Despite such difficulties, however, we have seen that the
Born approximation in its present form provides a good fit to
numerical scattering cross sections when $\lambda \gg r_s$, and it
remains a useful tool.

One intriguing question raised by this work is: when, if ever, are
fermion diffraction patterns from black holes of physical relevance?
Since diffraction effects are only significant when $r_s \sim
\lambda$, the question may be rephrased: do small black holes
exist in nature? By small, we mean black holes with an event
horizon smaller than the Compton wavelength of the lightest fermion
(thought to be the electron neutrino). Recent experimental evidence
suggests that neutrinos have non-zero mass. A neutrino with a
hypothetical mass $m_{\nu_e} \sim 0.01$ eV has a Compton wavelength of
$1 \times 10^{-5}$ m, and this size corresponds to a black hole of
mass $M \sim 10^{22}$ kg. Black holes of this magnitude or smaller are
unlikely to be created by astrophysical processes; however, it is
possible that small primordial black holes were created in the
early universe. An alternative scenario for small black hole genesis
arises in recently-proposed theories with Large Extra Dimensions
\cite{Arkani-Hamed-1998}. If such theories are well-founded, it is
possible that black holes may one day be created at a particle
accelerator (see \cite{Kanti-2004} for a review of these ideas).

Despite these possibilities, fermion diffraction patterns may never be
observed. A more physically-relevant problem is that of
gravitational-wave scattering from rotating, large black holes ($M
> 10^{30}$ kg). Previous studies of massless spin-two waves on the
Kerr spacetime have shown that the situation is complicated by effects
including frame-dragging, spin-rotation coupling, and
superradiance. Futterman, Handler and Matzner
\cite{Futterman-1988} present $s = 2$ scattering cross sections for a
special case: the alignment of black hole spin axis and incident wave
momentum. However, we are not aware of a general analysis of off-axis
gravitational scattering, though progress has been made for the
simpler spin-zero case \cite{Glampedakis-2001}. We suggest that a
general study of spin-half scattering from a Kerr black hole would be
a useful addition to the literature, for two reasons. Firstly, Unruh
\cite{Unruh-1973} has shown that superradiance does not occur for
massless spin-half waves. Secondly, though we would expect to see the effects
of spin-rotation coupling, the fewer degrees of freedom in
the $s=1/2$ case will make the results easier to interpret. Such a
study would contribute to our understanding of gravitational-wave
scattering, as well as being interesting in its own right. We
therefore hope to extend the methods of this paper to the Kerr black
hole in the near future.

\appendix

\section{Near-Horizon Scattering Approximation}
In \cite{Darwin-1959}, Darwin derived an approximation for the deflection $\Theta$ of an unbound massless particle passing close to the horizon. Here we extend that calculation to the massive case, where $v < 1$.

Following the approach of Chandrasekhar \cite{Chandrasekhar-1983}, we
look for roots of the orbit equation (\ref{classical-de}). For unbound
orbits that avoid the singularity, the orbit equation has three three
real roots, $u_1 < 0$, $u_3 > u_2 > 0$. The roots can be written in
terms of the eccentricity $e$ and latus rectum $l$
\begin{equation}
u_1 = -\frac{e - 1}{l},  \quad
u_2 = \frac{e+1}{l},  \quad
u_3 = \frac{1}{2M} - \frac{2}{l} 
\end{equation}
where $e$ and $l$ satisfy
\begin{align}
\frac{1 - \mu (3 + e^2)}{M l} = \frac{1 - v^2}{v^2 b^2} \, , \nn \\
\frac{1}{l^2} (e^2 - 1)(1 - 4\mu) = \frac{1}{b^2} \, , 
\end{align}
and $\mu = M / l$. Making the substitution $u = (1 + e \cos \chi) / l$ in (\ref{classical-de}) yields
\begin{equation}
\frac{d \chi}{d \phi} = (1 - 6\mu + 2\mu e)^{1/2} \left[1 - k^2 \cos^2(\chi / 2) \right]^{1/2}
\label{elliptic-scattering-eq}
\end{equation}
where 
\begin{equation}
k^2 = \frac{4 \mu e}{1 - 6 \mu + 2\mu e} 
\end{equation}
The variable $\chi$ runs from $0 \leq \chi \leq \chi_\infty = \cos^{-1} (-e^{-1})$. The solutions of (\ref{elliptic-scattering-eq}) are elliptic functions, and we can write the final angle as
\begin{equation}
\phi_\infty = \frac{2}{(1 - 6\mu + 2\mu e)^{1/2}} \left[ K(k) - F(\half \pi - \half \chi_\infty, k) \right]
\label{defleq}
\end{equation}
The critical orbit occurs where $u_2 = u_3$, $k = 1$. Then
\begin{align}
e_c &= \sqrt{1 + 8v^2} \, , \nn \\
l_c &= 2M (\sqrt{1 + 8v^2} + 3) \, , \nn \\
b_c &= \frac{\sqrt{8v^4 + 20v^2 - 1 + (1 + 8v^2)^{3/2}}M}{\sqrt{2} v^2}  .
\end{align}
The critical parameters are functions of the particle speed $v$ only and lie in the ranges $1 \leq e_c \leq 3$, $8M \leq l_c \leq 12M$, and $3M \leq r_\text{min} \leq 4M$. We now consider orbits close to the critical orbit, by expanding in a power series in $\delta = l - l_c$
\begin{align}
l &= l_c + \delta \, , \nn \\
e &= e_c + \frac{4 v^2}{(e_c + 1)(e_c + 3)} \frac{\delta}{M} \, , \nn \\
b &= b_c + M \frac{e_c}{2 \sqrt{2} v (e_c + 3)^{1/2} (e_c + 1)^{5/2}} \frac{\delta^2}{M^2} \, \nn \\
k^2 &= 1 - \frac{1}{e_c(e_c+3)} \frac{\delta}{M} .
\end{align}
Note that $b$ varies only quadratically with $\delta$. We employ the following approximations for the elliptic functions
\begin{align}
K(k) &\approx \ln \frac{4}{(1 - k^2)^{1/2}} \, , \nn \\
F(z, 1) &\approx \half \ln \left( \frac{1 + \sin{z}}{1 - \sin{z}} \right) \, ,
\end{align}
and substitute into (\ref{defleq}). Finally, we define the deflection $\Theta = 2 \phi_\infty - \pi$ to obtain result (\ref{darwin2}).

\section{Second-order Born Series}
Doran and Lasenby \cite{Doran-2002} showed how to calculate the
first-order contribution to the fermion cross section in a consistent
way. By expanding the scattering amplitude as a power series in the
black hole mass $M$, they found the first term in the scattering
amplitude series to be
\begin{equation}
\Mcal_{1} \equiv \us \hat{\Gamma}_1 \ur \quad \text{,    where  }  \hat{\Gamma}_1 = -\frac{4 \pi GM}{|\bp_f - \bp_i|^2} (2E \gam^0 - m) 
\end{equation}
and demonstrated its gauge invariance. Below, we calculate the second term in the scattering
amplitude, $\Mcal_2$, in the Kerr-Schild gauge. We find the real part to be 
\begin{equation}
\Mcal_{2} = - \frac{\pi^2 G^2 M^2}{|\bp_f - \bp_i|} \us \left[(4E\gam_0 - m) + \frac{4 (E\gam_0 - m) (4E^2 - m^2)}{|\bp_i + \bp_f|^2}\left(1 - |\sin(\theta/2)| \right)\right] \ur 
\end{equation}
which yields a second-order contribution to the cross section given by
(\ref{second-order-scat}). As in the Coulomb case, the imaginary part
of the scattering amplitude contains a divergent term, but the resulting
polarization is finite at first-order.
%\begin{equation}
%{\frac{d \sig}{d \Omega}}_{(2)} = \frac{\pi G^3 M^3 E}{4 v^3
%\sin^3(\theta / 2)} \left( (3+4v^2+v^4)(2 - |\sin{\theta / 2}|) -
%(7v^2 + v^4) \sin^2 (\theta/2) \right)
%\end{equation}

The second order scattering amplitude, $\Mcal_2 = \us \hat{\Gamma}_2 \ur$, is defined by
\begin{equation}
 \hat{\Gamma}_2 =  \int \frac{d^3k}{(2\pi)^3} B(\bp_f, \bk) \frac{\ksl + m}{k^2 - m^2 + i\eps} B(\bk, \bp_i) ,
\label{Gammatwo}
\end{equation}
where
\begin{align}
B(\bp_2, \bp_1) &= - \frac{2\pi GM}{|\bq|^2} (4E\gam_0 - \psl_1 - \psl_2)  
                  - \frac{4\pi GM}{|\bq|^4} ({\bp_2}^2 - {\bp_1}^2)(\psl_2 - \psl_1) \nn \\
  &               + \frac{i \pi^2 G M}{|\bq|^3} \left[ ({\bp_2}^2 - {\bp_1}^2) \gam_0 - 2E ( \psl_2 - \psl_1) \right] ,
\end{align}
and $\bq = \bp_2 - \bp_1$. Our task is made easier by splitting
the calculation into several parts, so that
\begin{equation}
B(\bp_2, \bp_1) = c_1 \hat{T}_1(\bp_2, \bp_1) + c_2 \hat{T}_2(\bp_2, \bp_1) + i c_3 \hat{T}_3(\bp_2, \bp_1) ,
\end{equation}
where
\begin{align}
(1) \quad & c_1 = - 2\pi GM,  &  \hat{T}_1(\bp_2, \bp_1) &= \frac{ (4E\gam_0 - \psl_1 - \psl_2) }{|\bq|^2}, \nn \\
(2) \quad & c_2 = - 4\pi GM,  &  \hat{T}_2(\bp_2, \bp_1) &= \frac{({\bp_2}^2 - {\bp_1}^2)(\psl_2 - \psl_1)} {|\bq|^4}, \nn \\
(3) \quad & i c_3 = i \pi^2 GM, &  \hat{T}_3(\bp_2, \bp_1) &= \frac{({\bp_2}^2 - {\bp_1}^2)\gam_0 - 2E(\psl_2 - \psl_1) }{|\bq|^3} .
\end{align}
We now consider each pair of terms in (\ref{Gammatwo}) in turn, and wherever possible
employ the simplification that $\psl_i \ur = m \ur$ and
$\us \psl_f = \us m$.
% Also introduce the useful substitution $\rsl =(\psl_1 +\psl_2 ) / 2$, and simplify where possible with $\overline{u}_s(\bp_2) \rsl u_r(\bp_1) = \overline{u}_s(\bp_2) m u_r(\bp_1) $.

First, let us consider term (2 x 2), to illustrate a method for evaluating these integrals. 
\begin{align}
\text{(2 x 2)} &= {c_2}^2 \integrand \frac{(\bk^2 - \bp^2)^2 (\ksl - m)}{|\bk - \bp_f|^4 |\bk - \bp_i|^4} .
\end{align}
We start by defining $\bQ = (\bp_f - \bp_i)/2$, $\br = (\bp_i + \bp_f) / 2$ and moving to the centre of mass frame, where the 1 axis is along $\br$ and the 3 axis is along $\bQ$. Then
\begin{align}
\bk            &\mapsto  \bk + \br \nn  , \\
\ksl           &\mapsto  \ksl + \rsl - E\gam_0 \nn , \\
\bk^2 - \bp^2  &\mapsto  \bk^2 - \bQ^2 + 2 \br \cdot \bk  .
\end{align}
Next, we employ spheroidal coordinates, $\{ u , v, \phi \}$, so
\begin{align}
k_1 &= |\bQ| \sinh{u} \sin{v} \cos{\phi} \nn , \\
k_2 &= |\bQ| \sinh{u} \sin{v} \sin{\phi} \nn ,\\
k_3 &= |\bQ| \cosh{u} \cos{v} .
\end{align}
With these coordinates the measure of integration is
\begin{equation}
d^3 k = |\bQ|^3 \sinh{u} \sin{v} \left( \sinh^2{u} + \sin^2{v} \right) du dv d\phi 
\end{equation}
and the important quantities in the integral become
\begin{align}
|\bk - \bp_2| |\bk - \bp_1 | &\mapsto |\bk - \bQ| |\bk + \bQ| = |\bQ|^2 (\sinh^2 {u} + \sin^2 {v} ) \nn \\
\bk^2 - \bQ^2 &= |\bQ|^2 ( \sinh^2{u} - \sin^2{v} ) .
\end{align}
With these replacements, $\ksl - m \mapsto \ksl + \rsl - E \gam_0 - m = \ksl - E \gam_0$. This has only a spatial component, so will couple to the odd part $\br \cdot \bk$ of the rest of the integral. Therefore
\begin{align}
\text{(2 x 2)} &= {c_2}^2 \integrand \frac{4 (\bk^2 - \bQ^2) (\br \cdot \bk) (\ksl - E\gam_0)}{|\bk - \bQ|^4 |\bk + \bQ|^4} \nn \\
 &= \frac{{c_2}^2  (m - E \gam_0)}{2 |\bQ| \pi^2} \int du dv \frac{\sinh^3 u \sin^3 v (\sinh^2 u - \sin^2 v)}{(\sinh^2 u + \sin^2 v )^3} .
\end{align}
The integral evaluates to $\pi^2 / 16$, and the result is
\begin{equation}
\text{(2 x 2)} = \frac{\pi^2 G^2 M^2}{|\bp_f - \bp_i|} (m - E \gam_0) .
\label{2by2result}
\end{equation}
Similar techniques can be used to evaluate terms (3 x 3), and (1 x 2) + (2 x 1). We find
\begin{align}
\text{(3 x 3)}           &= \frac{\pi^2 G^2 M^2}{2 |\bp_f - \bp_i|}  (m - 7 E \gam_0) , \label{3by3result} \\
\text{(1 x 2) + (2 x 1)} &= \frac{\pi^2 G^2 M^2}{2 |\bp_f - \bp_i|}  (9 E \gam_0 - 5 m) \label{1by2result} .
\end{align}
The other cross terms, (1 x 3) and (2 x 3), can be shown to be zero by symmetry considerations. The remaining term, (1 x 1), has two parts: an integral which is tractable using the above techniques, and an integral with a pole at $|\bk| = |\bp|$ that is familiar from the second-order Coulomb calculation of Dalitz \cite{Dalitz-1951}. Explicitly,
\begin{align}
\text{(1 x 1)} = \left( 2 \pi^2 G^2 M^2 \right) \left[ \frac{(m - 2 E \gam_0)}{|\bp_f - \bp_i|} \right. & + (2E\gam_0-m)^2(m+E\gam_0) (I / \pi^3) \nn \\ & \left. + (4E^2 - m^2) (E\gam_0 - m) (J / \pi^3) \right]
\label{1by1result}
\end{align}
where $I$ and $J$ are defined in terms of a low-frequency cut-off $\lambda$ as 
\begin{align}
I &= \int \frac{d^3 k}{[(\bk - \bp_f)^2 + \lam^2][(\bk - \bp_i)^2 + \lam^2](\ksl^2 - \psl^2 + i \eps)} \nn \\
\frac{(\psl_i + \psl_f)_r}{2} J &= \int \frac{k_r d^3 k}{[(\bk - \bp_f)^2 + \lam^2][(\bk - \bp_i)^2 + \lam^2](\ksl^2 - \psl^2 + i \eps)} .
\end{align}
To lowest order in $\lam$ these integrals evaluate to (see
 Itzykson and Zuber \cite{Itzykson-1980})
\begin{align}
I &= \frac{-i \pi^2}{2 p^3 \sin^2(\theta / 2)} \ln \left(\frac{2 p |\sin(\theta/2)|}{\lam} \right) \nn \\
J &= \sec^2(\theta / 2) I + \frac{\pi^3}{4 p^3 \cos^2(\theta / 2)} \left( 1 - \csc(\theta / 2) \right) - \frac{i \pi^2}{2p^3 \cos^2(\theta / 2)} \ln(\lam / 2p) \nn \\
I - J &= \frac{\pi^3}{4 p^3 \cos^2(\theta / 2)} \left( \csc(\theta / 2) - 1 \right) + \frac{i \pi^2}{4 p^3 \cos^2 (\theta/2)} \ln(\sin^2(\theta / 2)) .
\end{align}
The imaginary parts of $I$ and $J$ diverge as $\lam \rightarrow 0$. However we note that the difference $I - J$ is finite. We can rewrite the $I$ and $J$ parts of (1 x 1) as
\begin{equation}
2 I (2E^2 - m^2)(2E\gam_0 - m) + (I - J) (4E^2 - m^2) (m - E\gam_0) .
\end{equation}
The $2E \gam_0 - m$ associated with the $I$ term ensures that this term does not contribute to the polarization at first order, as we see below.

The result of summing (\ref{2by2result}), (\ref{3by3result}), (\ref{1by2result}) and (\ref{1by1result}) is
\begin{align}
\hat{\Gamma}_{2} = &- \frac{\pi^2 G^2 M^2}{|\bp_f - \bp_i|} \left[(4E\gam_0 - m) + \frac{4 (E\gam_0 - m) (4E^2 - m^2)}{|\bp_i + \bp_f|^2}\left(1 - |\sin(\theta/2)| \right)\right] \nn \\
& + i \frac{\pi G^2 M^2 (4E^2 - m^2) \ln(\sin^2(\theta/2))}{2 p^3 \cos^2(\theta/2)}  (m - E\gam_0) + i \infty (2E \gam_0 - m) .
\label{Gamma-2}
\end{align}

The second order contribution to the scattering cross section is
\begin{equation}
\left( \frac{d \sig}{d \Omega} \right)_2 = \left(\frac{m}{2\pi}\right)^2 \frac{1}{2} \left( \Tr \left\{ \hat{\overline{\Gamma}}_1 \frac{\psl_i + m}{2m} \hat{\Gamma}_2 \frac{\psl_f + m}{2m} \right\} + \Tr \left\{ \hat{\overline{\Gamma}}_2 \frac{\psl_i + m}{2m} \hat{\Gamma}_1 \frac{\psl_f + m}{2m} \right\}  \right) ,
\end{equation}
where $\hat{\overline{\Gamma}} = \gam_0 \hat{\Gamma}^\dagger \gam_0$. After
employing the appropriate gamma matrix trace theorems, this evaluates
to (\ref{second-order-scat}). The imaginary part of (\ref{Gamma-2}) does
not contribute to the total unpolarized scattering cross section, but
it does contribute to the net polarization. The polarized cross
section in direction $\hat{\bs}$ is given by
\begin{align}
\left( \frac{d \sig}{d \Omega} \right)_\text{pol} &= \left(\frac{1}{2\pi}\right)^2 \frac{1}{16} \left( \Tr \left\{ \gam_5 \ssl \hat{\overline{\Gamma}}_1 \psl_i \hat{\Gamma}_2 \psl_f \right\}  +
  \Tr \left\{ \gam_5 \ssl \hat{\overline{\Gamma}}_2 \psl_i \hat{\Gamma}_1 \psl_f \right\}  \right) \nn \\ 
 &= \frac{2 G^3 M^3 E \mass}{\pi^2 |\bp_f - \bp_i|^2} (4E^2 - m^2) \text{Im}( I - J ) \hat{\bs} \cdot (\bp_i \times \bp_f) \label{pol-abs}.
\end{align}
Here we have made use of the result
\begin{equation}
\Tr \left\{ \gam_5 \asl \bsl \csl \dsl \right\} = -4 i \eps_{\alpha \beta \gamma \delta} a^\alpha b^\beta c^\gamma d^\delta ,
\end{equation}
where $\eps_{\alpha \beta \gamma \delta}$ is antisymmetric under exchange of any pair of indices. The fraction of polarized flux $\polP$ is found by dividing polarized flux (\ref{pol-abs}) by the unpolarized flux at first order (\ref{first-order-scat}) to obtain result (\ref{second-order-pol}).

\bibliographystyle{apsrev}

\begin{thebibliography}{41}
\expandafter\ifx\csname natexlab\endcsname\relax\def\natexlab#1{#1}\fi
\expandafter\ifx\csname bibnamefont\endcsname\relax
  \def\bibnamefont#1{#1}\fi
\expandafter\ifx\csname bibfnamefont\endcsname\relax
  \def\bibfnamefont#1{#1}\fi
\expandafter\ifx\csname citenamefont\endcsname\relax
  \def\citenamefont#1{#1}\fi
\expandafter\ifx\csname url\endcsname\relax
  \def\url#1{\texttt{#1}}\fi
\expandafter\ifx\csname urlprefix\endcsname\relax\def\urlprefix{URL }\fi
\providecommand{\bibinfo}[2]{#2}
\providecommand{\eprint}[2][]{\url{#2}}

\bibitem[{\citenamefont{Matzner}(1968)}]{Matzner-1968}
\bibinfo{author}{\bibfnamefont{R.~A.} \bibnamefont{Matzner}},
  \bibinfo{journal}{J. Math. Phys.} \textbf{\bibinfo{volume}{9}},
  \bibinfo{pages}{163} (\bibinfo{year}{1968}).

\bibitem[{\citenamefont{Fabbri}(1975)}]{Fabbri-1975}
\bibinfo{author}{\bibfnamefont{R.}~\bibnamefont{Fabbri}},
  \bibinfo{journal}{Phys. Rev. D} \textbf{\bibinfo{volume}{12}},
  \bibinfo{pages}{933} (\bibinfo{year}{1975}).

\bibitem[{\citenamefont{Peters}(1976)}]{Peters-1976}
\bibinfo{author}{\bibfnamefont{P.~C.} \bibnamefont{Peters}},
  \bibinfo{journal}{Phys. Rev. D} \textbf{\bibinfo{volume}{13}},
  \bibinfo{pages}{775} (\bibinfo{year}{1976}).

\bibitem[{\citenamefont{de~Logi and S.~J.~Kov\'acs}(1977)}]{DeLogi-1977}
\bibinfo{author}{\bibfnamefont{W.~K.} \bibnamefont{de~Logi}} \bibnamefont{and}
  \bibinfo{author}{\bibfnamefont{S.~J.}~\bibnamefont{Kov\'acs}},
  \bibinfo{journal}{Phys. Rev. D} \textbf{\bibinfo{volume}{16}},
  \bibinfo{pages}{237} (\bibinfo{year}{1977}).

\bibitem[{\citenamefont{S\'anchez}(1976)}]{Sanchez-1976}
\bibinfo{author}{\bibfnamefont{N.~G.} \bibnamefont{S\'anchez}},
  \bibinfo{journal}{J. Math. Phys.} \textbf{\bibinfo{volume}{17}},
  \bibinfo{pages}{688} (\bibinfo{year}{1976}).

\bibitem[{\citenamefont{S\'anchez}(1977)}]{Sanchez-1977}
\bibinfo{author}{\bibfnamefont{N.~G.} \bibnamefont{S\'anchez}},
  \bibinfo{journal}{Phys. Rev. D} \textbf{\bibinfo{volume}{16}},
  \bibinfo{pages}{937} (\bibinfo{year}{1977}).

\bibitem[{\citenamefont{S\'anchez}(1978{\natexlab{a}})}]{Sanchez-1978a}
\bibinfo{author}{\bibfnamefont{N.~G.} \bibnamefont{S\'anchez}},
  \bibinfo{journal}{Phys. Rev. D} \textbf{\bibinfo{volume}{18}},
  \bibinfo{pages}{1030} (\bibinfo{year}{1978}{\natexlab{a}}).

\bibitem[{\citenamefont{S\'anchez}(1978{\natexlab{b}})}]{Sanchez-1978b}
\bibinfo{author}{\bibfnamefont{N.~G.} \bibnamefont{S\'anchez}},
  \bibinfo{journal}{Phys. Rev. D} \textbf{\bibinfo{volume}{18}},
  \bibinfo{pages}{1798} (\bibinfo{year}{1978}{\natexlab{b}}).

\bibitem[{\citenamefont{Zhang and DeWitt-Morette}(1984)}]{Zhang-1984}
\bibinfo{author}{\bibfnamefont{T.-R.} \bibnamefont{Zhang}} \bibnamefont{and}
  \bibinfo{author}{\bibfnamefont{C.}~\bibnamefont{DeWitt-Morette}},
  \bibinfo{journal}{Phys. Rev. Lett.} \textbf{\bibinfo{volume}{52}},
  \bibinfo{pages}{2313} (\bibinfo{year}{1984}).

\bibitem[{\citenamefont{Matzner et~al.}(1985)\citenamefont{Matzner,
  DeWitt-Morette, Nelson, and Zhang}}]{Matzner-1985}
\bibinfo{author}{\bibfnamefont{R.~A.} \bibnamefont{Matzner}},
  \bibinfo{author}{\bibfnamefont{C.}~\bibnamefont{DeWitt-Morette}},
  \bibinfo{author}{\bibfnamefont{B.}~\bibnamefont{Nelson}}, \bibnamefont{and}
  \bibinfo{author}{\bibfnamefont{T.-R.} \bibnamefont{Zhang}},
  \bibinfo{journal}{Phys. Rev. D} \textbf{\bibinfo{volume}{31}},
  \bibinfo{pages}{1869} (\bibinfo{year}{1985}).

\bibitem[{\citenamefont{Anninos et~al.}(1992)\citenamefont{Anninos,
  DeWitt-Morette, Matzner, Yioutas, and Zhang}}]{Anninos-1992}
\bibinfo{author}{\bibfnamefont{P.}~\bibnamefont{Anninos}},
  \bibinfo{author}{\bibfnamefont{C.}~\bibnamefont{DeWitt-Morette}},
  \bibinfo{author}{\bibfnamefont{R.~A.} \bibnamefont{Matzner}},
  \bibinfo{author}{\bibfnamefont{P.}~\bibnamefont{Yioutas}}, \bibnamefont{and}
  \bibinfo{author}{\bibfnamefont{T.~R.} \bibnamefont{Zhang}},
  \bibinfo{journal}{Phys. Rev. D} \textbf{\bibinfo{volume}{46}},
  \bibinfo{pages}{4477} (\bibinfo{year}{1992}).

\bibitem[{\citenamefont{Andersson}(1995)}]{Andersson-1995}
\bibinfo{author}{\bibfnamefont{N.}~\bibnamefont{Andersson}},
  \bibinfo{journal}{Phys. Rev. D} \textbf{\bibinfo{volume}{52}},
  \bibinfo{pages}{1808} (\bibinfo{year}{1995}).

\bibitem[{\citenamefont{Andersson and Jensen}(2001)}]{Andersson-2001}
\bibinfo{author}{\bibfnamefont{N.}~\bibnamefont{Andersson}} \bibnamefont{and}
  \bibinfo{author}{\bibfnamefont{B.~P.} \bibnamefont{Jensen}},
  \bibinfo{journal}{arxiv}  (\bibinfo{year}{2001}), \eprint{gr-qc/0011025}.

\bibitem[{\citenamefont{Doran and Lasenby}(2002)}]{Doran-2002}
\bibinfo{author}{\bibfnamefont{C.~J.~L.} \bibnamefont{Doran}} \bibnamefont{and}
  \bibinfo{author}{\bibfnamefont{A.~N.} \bibnamefont{Lasenby}},
  \bibinfo{journal}{Phys. Rev. D} \textbf{\bibinfo{volume}{66}},
  \bibinfo{pages}{024006} (\bibinfo{year}{2002}).

\bibitem[{\citenamefont{Chandrasekhar}(1983)}]{Chandrasekhar-1983}
\bibinfo{author}{\bibfnamefont{S.}~\bibnamefont{Chandrasekhar}},
  \emph{\bibinfo{title}{The Mathematical Theory of Black Holes}}, International
  Series of Monographs on Physics (\bibinfo{publisher}{Oxford University Press, New York}, \bibinfo{year}{1983}).

\bibitem[{\citenamefont{Futterman et~al.}(1988)\citenamefont{Futterman,
  Handler, and Matzner}}]{Futterman-1988}
\bibinfo{author}{\bibfnamefont{J.~A.~H.} \bibnamefont{Futterman}},
  \bibinfo{author}{\bibfnamefont{F.~A.} \bibnamefont{Handler}},
  \bibnamefont{and} \bibinfo{author}{\bibfnamefont{R.~A.}
  \bibnamefont{Matzner}}, \emph{\bibinfo{title}{Scattering from Black Holes}}
  (\bibinfo{publisher}{Cambridge University Press}, \bibinfo{year}{1988}).

\bibitem[{\citenamefont{Frolov and Novikov}(1998)}]{Frolov-1998}
\bibinfo{author}{\bibfnamefont{V.~P.} \bibnamefont{Frolov}} \bibnamefont{and}
  \bibinfo{author}{\bibfnamefont{I.~D.} \bibnamefont{Novikov}},
  \emph{\bibinfo{title}{Black hole physics: Basic concepts and new
  developments}} (\bibinfo{publisher}{Kluwer Academic Publishers, Dordrecht},
  \bibinfo{year}{1998}).

\bibitem[{\citenamefont{Unruh}(1976)}]{Unruh-1976-absorption}
\bibinfo{author}{\bibfnamefont{W.~G.} \bibnamefont{Unruh}},
  \bibinfo{journal}{Phys. Rev. D} \textbf{\bibinfo{volume}{14}},
  \bibinfo{pages}{3251} (\bibinfo{year}{1976}).

\bibitem[{\citenamefont{Darwin}(1959)}]{Darwin-1959}
\bibinfo{author}{\bibfnamefont{C.}~\bibnamefont{Darwin}},
  \bibinfo{journal}{Proc. R. Soc. Lond. A} \textbf{\bibinfo{volume}{249}},
  \bibinfo{pages}{180} (\bibinfo{year}{1959}).

\bibitem[{\citenamefont{Collins et~al.}(1973)\citenamefont{Collins, Delbourgo,
  and Williams}}]{Collins-1973}
\bibinfo{author}{\bibfnamefont{P.~A.} \bibnamefont{Collins}},
  \bibinfo{author}{\bibfnamefont{R.}~\bibnamefont{Delbourgo}},
  \bibnamefont{and} \bibinfo{author}{\bibfnamefont{R.~M.}
  \bibnamefont{Williams}}, \bibinfo{journal}{J. Phys. A.}
  \textbf{\bibinfo{volume}{6}}, \bibinfo{pages}{161} (\bibinfo{year}{1973}).

\bibitem[{\citenamefont{Dalitz}(1951)}]{Dalitz-1951}
\bibinfo{author}{\bibfnamefont{R.~H.} \bibnamefont{Dalitz}},
  \bibinfo{journal}{Royal Society of London Proceedings Series A}
  \textbf{\bibinfo{volume}{206}}, \bibinfo{pages}{509} (\bibinfo{year}{1951}).

\bibitem[{\citenamefont{Ford and Wheeler}(1959)}]{Ford-1959}
\bibinfo{author}{\bibfnamefont{K.~W.} \bibnamefont{Ford}} \bibnamefont{and}
  \bibinfo{author}{\bibfnamefont{J.~A.} \bibnamefont{Wheeler}},
  \bibinfo{journal}{Annals of Physics} \textbf{\bibinfo{volume}{7}},
  \bibinfo{pages}{259} (\bibinfo{year}{1959}).

\bibitem[{\citenamefont{Rose}(1961)}]{Rose-1961}
\bibinfo{author}{\bibfnamefont{M.}~\bibnamefont{Rose}},
  \emph{\bibinfo{title}{Relativistic Electron Theory}}
  (\bibinfo{publisher}{John Wiley \& Sons}, \bibinfo{year}{1961}).

\bibitem[{\citenamefont{Mott and Massey}(1965)}]{Mott-1965}
\bibinfo{author}{\bibfnamefont{N.~F.} \bibnamefont{Mott}} \bibnamefont{and}
  \bibinfo{author}{\bibfnamefont{H.~S.~W.} \bibnamefont{Massey}},
  \emph{\bibinfo{title}{The Theory of Atomic Collisions}}
  (\bibinfo{publisher}{Oxford University Press, London}, \bibinfo{year}{1965}).

\bibitem[{\citenamefont{Martel and Poisson}(2001)}]{Martel-2001}
\bibinfo{author}{\bibfnamefont{K.}~\bibnamefont{Martel}} \bibnamefont{and}
  \bibinfo{author}{\bibfnamefont{E.}~\bibnamefont{Poisson}},
  \bibinfo{journal}{Am. J. Phys.} \textbf{\bibinfo{volume}{69}},
  \bibinfo{pages}{476} (\bibinfo{year}{2001}).

\bibitem[{\citenamefont{Birrell and Davies}(1982)}]{Birrell-1982}
\bibinfo{author}{\bibfnamefont{N.~D.} \bibnamefont{Birrell}} \bibnamefont{and}
  \bibinfo{author}{\bibfnamefont{P.~C.~W.} \bibnamefont{Davies}},
  \emph{\bibinfo{title}{Quantum Fields in Curved Space}}
  (\bibinfo{publisher}{Cambridge University Press}, \bibinfo{year}{1982}).

\bibitem[{\citenamefont{Lasenby et~al.}(2005)\citenamefont{Lasenby, Doran,
  Pritchard, Caceres, and Dolan}}]{Lasenby-2005-bs}
\bibinfo{author}{\bibfnamefont{A.~N.} \bibnamefont{Lasenby}},
  \bibinfo{author}{\bibfnamefont{C.~J.~L.} \bibnamefont{Doran}},
  \bibinfo{author}{\bibfnamefont{J.}~\bibnamefont{Pritchard}},
  \bibinfo{author}{\bibfnamefont{A.}~\bibnamefont{Caceres}}, \bibnamefont{and}
  \bibinfo{author}{\bibfnamefont{S.~R.} \bibnamefont{Dolan}},
  \bibinfo{journal}{Phys. Rev. D} \textbf{\bibinfo{volume}{72}},
  \bibinfo{pages}{105014} (\bibinfo{year}{2005}).

\bibitem[{\citenamefont{Nakahara}(1990)}]{Nakahara-1990}
\bibinfo{author}{\bibfnamefont{M.}~\bibnamefont{Nakahara}},
  \emph{\bibinfo{title}{Geometry, Topology and Physics}}
  (\bibinfo{publisher}{Adam Hilger, Bristol}, \bibinfo{year}{1990}).

\bibitem[{\citenamefont{Leaver}(1986)}]{Leaver-1986}
\bibinfo{author}{\bibfnamefont{E.~W.} \bibnamefont{Leaver}},
  \bibinfo{journal}{J. Math. Phys.} \textbf{\bibinfo{volume}{27}},
  \bibinfo{pages}{1238} (\bibinfo{year}{1986}).

\bibitem[{\citenamefont{Mano et~al.}(1996)\citenamefont{Mano, Suzuki, and
  Takasugi}}]{Mano-1996}
\bibinfo{author}{\bibfnamefont{S.}~\bibnamefont{Mano}},
  \bibinfo{author}{\bibfnamefont{H.}~\bibnamefont{Suzuki}}, \bibnamefont{and}
  \bibinfo{author}{\bibfnamefont{E.}~\bibnamefont{Takasugi}},
  \bibinfo{journal}{Prog. Theor. Phys.} \textbf{\bibinfo{volume}{96}},
  \bibinfo{pages}{549} (\bibinfo{year}{1996}).

\bibitem[{\citenamefont{Lasenby et~al.}(1998)\citenamefont{Lasenby, Doran, and
  Gull}}]{Lasenby-1998-gtg}
\bibinfo{author}{\bibfnamefont{A.~N.} \bibnamefont{Lasenby}},
  \bibinfo{author}{\bibfnamefont{C.~J.~L.} \bibnamefont{Doran}},
  \bibnamefont{and} \bibinfo{author}{\bibfnamefont{S.~F.} \bibnamefont{Gull}},
  \bibinfo{journal}{Phil. Trans. R. Soc. Lond. A}
  \textbf{\bibinfo{volume}{356}}, \bibinfo{pages}{487} (\bibinfo{year}{1998}).

\bibitem[{\citenamefont{Andersson and Thylwe}(1994)}]{Andersson-1994}
\bibinfo{author}{\bibfnamefont{N.}~\bibnamefont{Andersson}} \bibnamefont{and}
  \bibinfo{author}{\bibfnamefont{K.-E.} \bibnamefont{Thylwe}},
  \bibinfo{journal}{Class. Quantum Grav.} \textbf{\bibinfo{volume}{11}},
  \bibinfo{pages}{2991} (\bibinfo{year}{1994}).

\bibitem[{\citenamefont{Wu and Ohmura}(1962)}]{Wu-1962}
\bibinfo{author}{\bibfnamefont{T.~Y.} \bibnamefont{Wu}} \bibnamefont{and}
  \bibinfo{author}{\bibfnamefont{T.}~\bibnamefont{Ohmura}},
  \emph{\bibinfo{title}{Quantum Theory of Scattering}}
  (\bibinfo{publisher}{Prentice Hall,}, \bibinfo{year}{1962}).

\bibitem[{\citenamefont{Yennie et~al.}(1954)\citenamefont{Yennie, Ravenhall,
  and Wilson}}]{Yennie-1954}
\bibinfo{author}{\bibfnamefont{D.~R.} \bibnamefont{Yennie}},
  \bibinfo{author}{\bibfnamefont{D.~G.} \bibnamefont{Ravenhall}},
  \bibnamefont{and} \bibinfo{author}{\bibfnamefont{R.~N.}
  \bibnamefont{Wilson}}, \bibinfo{journal}{Physical Review}
  \textbf{\bibinfo{volume}{95}}, \bibinfo{pages}{500} (\bibinfo{year}{1954}).

\bibitem[{\citenamefont{Starobinskii and Churilov}(1973)}]{Starobinskii-1973}
\bibinfo{author}{\bibfnamefont{A.~A.} \bibnamefont{Starobinskii}}
  \bibnamefont{and} \bibinfo{author}{\bibfnamefont{S.~M.}
  \bibnamefont{Churilov}}, \bibinfo{journal}{Zh. Eksp. Teor. Fiz.}
  \textbf{\bibinfo{volume}{65}}, \bibinfo{pages}{3} (\bibinfo{year}{1973}).

\bibitem[{\citenamefont{Doran et~al.}(2005)\citenamefont{Doran, Lasenby, Dolan,
  and Hinder}}]{Doran-2005-abs}
\bibinfo{author}{\bibfnamefont{C.~J.~L.} \bibnamefont{Doran}},
  \bibinfo{author}{\bibfnamefont{A.~N.} \bibnamefont{Lasenby}},
  \bibinfo{author}{\bibfnamefont{S.~R.} \bibnamefont{Dolan}}, \bibnamefont{and}
  \bibinfo{author}{\bibfnamefont{I.}~\bibnamefont{Hinder}},
  \bibinfo{journal}{Phys. Rev. D} \textbf{\bibinfo{volume}{71}},
  \bibinfo{pages}{124020} (\bibinfo{year}{2005}).

\bibitem[{\citenamefont{Arkani-Hamed et~al.}(1998)\citenamefont{Arkani-Hamed,
  Dimopoulos, and Dvali}}]{Arkani-Hamed-1998}
\bibinfo{author}{\bibfnamefont{N.}~\bibnamefont{Arkani-Hamed}},
  \bibinfo{author}{\bibfnamefont{S.}~\bibnamefont{Dimopoulos}},
  \bibnamefont{and} \bibinfo{author}{\bibfnamefont{G.}~\bibnamefont{Dvali}},
  \bibinfo{journal}{Phys. Lett. B} \textbf{\bibinfo{volume}{429}},
  \bibinfo{pages}{263} (\bibinfo{year}{1998}).

\bibitem[{\citenamefont{Kanti}(2004)}]{Kanti-2004}
\bibinfo{author}{\bibfnamefont{P.}~\bibnamefont{Kanti}},
  \bibinfo{journal}{Int. J. Mod. Phys. A} \textbf{\bibinfo{volume}{19}},
  \bibinfo{pages}{4899} (\bibinfo{year}{2004}).

\bibitem[{\citenamefont{Glampedakis and Andersson}(2001)}]{Glampedakis-2001}
\bibinfo{author}{\bibfnamefont{K.}~\bibnamefont{Glampedakis}} \bibnamefont{and}
  \bibinfo{author}{\bibfnamefont{N.}~\bibnamefont{Andersson}},
  \bibinfo{journal}{Class. Quantum Grav.} \textbf{\bibinfo{volume}{18}},
  \bibinfo{pages}{1939} (\bibinfo{year}{2001}).

\bibitem[{\citenamefont{Unruh}(1973)}]{Unruh-1973}
\bibinfo{author}{\bibfnamefont{W.}~\bibnamefont{Unruh}},
  \bibinfo{journal}{Phys. Rev. Lett.} \textbf{\bibinfo{volume}{31}},
  \bibinfo{pages}{1265} (\bibinfo{year}{1973}).

\bibitem[{\citenamefont{Itzykson and Zuber}(1980)}]{Itzykson-1980}
\bibinfo{author}{\bibfnamefont{C.}~\bibnamefont{Itzykson}} \bibnamefont{and}
  \bibinfo{author}{\bibfnamefont{J.-B.} \bibnamefont{Zuber}},
  \emph{\bibinfo{title}{Quantum Field Theory}}
  (\bibinfo{publisher}{McGraw-Hill}, \bibinfo{address}{New York},
  \bibinfo{year}{1980}).

\end{thebibliography}

\end{document}